\documentclass[aps,prb, twocolumn,floatfix,notitlepage, superscriptaddress,longbibliography]{revtex4-2}
\usepackage[utf8]{inputenc}
\usepackage{amsfonts,amsmath,amssymb,graphicx,epstopdf,verbatim}
\usepackage{mathbbol}
\usepackage{rotating}
\usepackage{epstopdf}
\usepackage{xcolor}
 
\usepackage[colorlinks]{hyperref}
\definecolor{darkred}{rgb}{0.5,0,0}
\definecolor{darkgreen}{rgb}{0,0.5,0}
\definecolor{darkblue}{rgb}{0,0,0.5}
\hypersetup{colorlinks,
linkcolor=darkred,
filecolor=darkgreen,
urlcolor=darkblue,
citecolor=darkgreen}
\usepackage{braket}
\usepackage{graphicx}
 
\usepackage{tikz}
\usetikzlibrary{shapes.geometric}
\usetikzlibrary{decorations.markings}
\usetikzlibrary{decorations.pathmorphing,patterns}
\usetikzlibrary{decorations}
\usetikzlibrary{calc}
\usetikzlibrary{intersections}
\usetikzlibrary{arrows}
\usetikzlibrary{matrix}
\usetikzlibrary{positioning}
\PassOptionsToPackage{dvipsnames,svgnames}{xcolor}     
\usepackage{xkeyval,tkz-base}
\usetikzlibrary{arrows,calc}

\newcommand{\blambda}{\boldsymbol{\lambda}}
\newcommand{\cycle}{\mathrm{cycle}}
\newcommand{\tauun}{\tau_\mathrm{un}}
\newcommand{\tauth}{\tau_\mathrm{th}}
\newcommand{\tauuntilde}{\tilde{\tau}_\mathrm{un}}
\newcommand{\Uforw}{U_f}
\newcommand{\Uback}{U_b}

\newcommand{\C}{{\mathbf C}}
\newcommand{\btheta}{\boldsymbol{\theta}}
\newcommand{\balpha}{\boldsymbol{\alpha}}

\newcommand{\Heis}{\mathrm{\scriptscriptstyle H}}

\newcommand{\transpose}{\mathsf{T}}

\newcommand{\Tr}{{\rm Tr}}

\newcommand{\res}{\mathrm{res}}
\newcommand{\lin}{\mathrm{lin}}
\newcommand{\opt}{\mathrm{opt}}
\newcommand{\thrs}{\scriptscriptstyle\mathrm{thrs}}
\newcommand{\threshold}{\epsilon^{\thrs}}
\newcommand{\eres}{e_{\res}}   %

\newcommand{\crit}{\mathrm{crit}}
\newcommand{\fast}{\mathrm{fast}}
\newcommand{\slow}{\mathrm{slow}}
\newcommand{\id}{\mathrm{id}}
\newcommand{\Power}{\mathcal{P}}
\newcommand{\opsym}[1]{#1}
\newcommand{\opc}[1]{{\opsym{c}^{\phantom{\dagger}}_{#1}}}
\newcommand{\opcdag}[1]{{\opsym{c}^{\dagger}_{#1}}}
\newcommand{\opb}[1]{{\opsym{b}^{\phantom{\dagger}}_{#1}}}
\newcommand{\opbdag}[1]{{\opsym{b}^{\dagger}_{#1}}}
 
\begin{document}
 
\title{From maximum work to maximum power in an optimally controlled quantum Ising Otto engine}
 
\author{Vincenzo Roberto Arezzo}
\affiliation{SISSA, Via Bonomea 265, I-34136 Trieste, Italy}
 
\author{Davide Rossini}
\affiliation{Dipartimento di Fisica dell’Universit\`a di Pisa and INFN, 
Largo Pontecorvo 3, I-56127 Pisa, Italy}
 
\author{Giuseppe E. Santoro}
 
\affiliation{SISSA, Via Bonomea 265, I-34136 Trieste, Italy}
\affiliation{International Centre for Theoretical Physics (ICTP), P.O. Box 586, I-34014 Trieste, Italy}
 
\author{Giulia Piccitto}
\affiliation{Dipartimento di Matematica e Informatica, Universit\`a di Catania, Viale Andrea Doria 6, I-95128 Catania, Italy}
 
\begin{abstract}
In finite-time quantum Otto engines, maximizing power requires balancing the
duration of the unitary strokes against the losses from nonadiabatic
excitations. We address this trade-off in an Otto engine with a free-fermion
working medium, using quantum optimal control. Having determined the
single-particle energy scales that maximize ideal work at fixed reservoir
temperatures, we realize them as the two endpoint Hamiltonians of a
transverse-field Ising chain, connected by a path that crosses the quantum
critical point. Optimal control of the unitary strokes yields near-ideal work
on a time scale growing linearly with system size, instead of quadratically as
for linear driving. We derive a time-rescaling relation that maps protocols
with different energy scales onto a single reference problem and, for complete
thermalization strokes of fixed duration, use it to build an analytical model
of the cycle power. The model predicts two competing maximum-power regimes: a
slow branch that approaches the ideal work, and a fast, size-independent branch
that trades work for a shorter cycle and dominates above a crossover size, in
quantitative agreement with numerical optimal-control results. It also
identifies power-optimal energy scales that differ from the work-optimal ones
and further increase the power. Our results provide a predictive approach to
jointly optimizing the operating conditions and finite-time dynamics of
quantum heat engines.
\end{abstract}
 
\maketitle

\section{Introduction}
\label{sec:intro}
Quantum thermodynamics aims at extending the concepts of heat, work, and
thermodynamic transformations to systems operating in the quantum regime.
A quantum heat engine converts heat absorbed from thermal reservoirs into
useful work through a suitable thermodynamic
cycle~\cite{alicki1979quantum, binder2019thermodynamics, deffner2019quantum, Kosloff_2014, Bhattacharjee_2021, Myers_2022}.  
Starting from the pioneering proposal of the maser as a quantum heat
engine~\cite{scovil1959three}, a variety of quantum thermodynamic cycles
have been investigated~\cite{quan2007quantum}, including Otto engines
based on few-level quantum systems~\cite{henrich2007quantum,Solfanelli_2020}.
Experimental realizations have also been demonstrated on different
platforms~\cite{rossnagel2016single, klatzow2019experimental, Abah_2012, peterson_2019, Zou_2017, Lindenfels_2019, maslennikov2019quantum, vanHorne_2020, Sheng_2021, Bouton_2021, deAssis_2021, Zhang_2022, Koch_2023, Uusnakki_2026}.
 
For an idealized engine made of independent two-level systems, the work
output grows extensively with the number of spins~\cite{quan2007quantum}
and, if the Hamiltonians at different times commute, the unitary strokes are
dynamically trivial, with no nonadiabatic losses~\cite{Rezek2006, Kosloff2017otto}.
In experimentally relevant spin chains, however, couplings and interactions
are generically present, and increasing the system size introduces not only
more available work, but also a growing dynamical cost, associated with
changing eigenbases, small gaps, and residual excitations~\cite{Dziarmaga_PRL2005,revathy2020universal}. Many-body working
substances have therefore attracted considerable attention~\cite{Mukherjee_2021,cangemi2023quantum,Jaramillo_2016,Chen_2019}, also because they open up the possibility of quantum critical behavior,
whose role in engine performance has been extensively
investigated~\cite{campisi2016power,Chen_2019, piccitto2022ising, Solfanelli_2023, Jaseem_2023, Williamson_2024, Xu_2024, Revathy_2024, Brollo_2025, Sajitha_2025, Katoorani_2025, Castorene_2026}: although criticality may serve as a resource
under specific conditions, its overall impact is often detrimental. It is therefore interesting to ask whether there exists an optimal way to implement thermodynamic transformations close to quantum critical points, in order to minimize dissipation effects. 

Nonadiabatic work losses can be avoided in the quasistatic limit of the unitary strokes, where the evolution is quantum adiabatic. This gives the maximum
extractable work, but no information on the power. For classical thermal engines operating in finite time, Curzon and Ahlborn~\cite{curzon1975efficiency} showed that an endoreversible Carnot engine with Newtonian heat transfer achieves an {\em efficiency at maximum power} of $\eta_{\mathrm{CA}}=1-\sqrt{T_c/T_h}$, below the Carnot efficiency $\eta_{\rm rev}=1-T_c/T_h$, where $T_c$ and $T_h$ are the cold and hot reservoir temperatures, respectively.
In the quantum case,
shortening the unitary strokes reduces the cycle duration but induces
nonadiabatic excitations that lower the work per cycle~\cite{Solfanelli_2020, del2022quantum, Kosloff_2002},
and the efficiency at maximum power of quantum Otto engines depends on how
these excitations are generated~\cite{Rezek2006,Abah_2012}. 
In particular, maximizing work through frictionless strokes does not generally maximize power, which can benefit from shorter strokes despite nonadiabatic losses~\cite{Rezek2006, Insinga_2020}. This
trade-off between work and power naturally motivates quantum-control
techniques to design faster unitary transformations~\cite{Caneva_PRL2009,Caneva_PRA2011,Doria_PRL2011,Muller_2022},
including shortcuts to adiabaticity for quantum heat engines~\cite{delCampo_2014}, their extension to many-body working
media~\cite{Beau_2016, Hartmann_2020}, and machine-learning optimization of the whole
cycle~\cite{Erdman_2022}.

In this work, we combine the optimal control of the schedule of the unitary
strokes with the optimization of the Hamiltonian energy scales for maximum
power, in a minimal, exactly solvable setting: a transverse-field Ising chain
driven across its quantum critical point between two endpoint Hamiltonians,
which is interacting in the spin representation but maps to free fermions. This allows us to retain a
nontrivial many-body control problem while performing systematic finite-size
scans.

The optimization of the engine involves two aspects. The first is
thermodynamic: we determine the Hamiltonian energy scales that maximize the
work output of an ideal Otto cycle with a free-fermion working medium. The
second is dynamical: keeping these work-optimal endpoints fixed, we optimize
the temporal schedule of the finite-time unitary strokes and find that, over the investigated system sizes, the
time needed to approach the ideal work grows only linearly with the system
size, instead of quadratically as for a linear ramp.
 
When the objective is the cycle power, however, the two problems cannot be
treated independently, since the nonadiabatic losses depend on both the
stroke duration and the energy scales: in principle, a separate
optimal-control problem would have to be solved for each choice of energy
scales and driving time. We avoid this by developing an analytical
description of the cycle power, based on a time-rescaling relation that
connects protocols with different Hamiltonian energy scales to a single
reference problem and, supported by the observed structure of the optimized
protocols, allows us to infer their nonadiabatic losses. The resulting model
predicts two competing maximum-power regimes: a slow branch, with longer
unitary strokes approaching the ideal work, and a fast branch, where shorter
strokes yield higher power despite reduced work, which dominates above a
crossover system size. It also determines the energy scales that maximize
power in the fast regime, which differ from those maximizing ideal work.
These predictions are quantitatively validated against numerical
optimal-control simulations, including endpoints and temperature ratios not
used to calibrate the model. Finally, we compare the efficiency at maximum
power of the two branches with the Curzon--Ahlborn value.
 
The paper is organized as follows. In Sec.~\ref{sec:The quantum Otto cycle}, we introduce the quantum Otto cycle and define the work and power output. In Sec.~\ref{sec:free_fermion_engine}, we determine the optimal ideal-work conditions for a free-fermion working medium, which are then realized in a transverse-field Ising chain in Sec.~\ref{sec:ising_engine}. In Sec.~\ref{subsec:ising_work_real}, we investigate finite-time work extraction using quantum optimal control. Finally, in Sec.~\ref{sec:power_optimization}, we develop the analytical
description of the cycle power and determine the optimal operating conditions.
Section~\ref{sec:conclusions} contains a summary and an outlook, while the
Appendices collect the more technical derivations.

\section{The quantum Otto cycle}
\label{sec:The quantum Otto cycle}
We focus on the quantum Otto cycle \cite{Kieu2004,Rezek2006,Kosloff2017otto} because it provides a particularly convenient framework for separating the coherent dynamics responsible for work extraction from the dissipative processes responsible for heat
exchange. In particular, its two unitary strokes allow us to formulate the
finite-time operation of the engine as a quantum-control problem, while the
thermalization strokes set the Hamiltonian eigenstate populations at the two
reservoir temperatures.
 
Let us consider a quantum system whose Hamiltonian explicitly depends on some external parameters $\blambda$ that can be varied in time, i.e. $H(t) = H(\blambda(t))$. The system is put in contact with two different reservoirs at temperatures $T_c$ and $T_h$, with $T_c < T_h$.
We assume that, when the system is in contact with the reservoir for a time $\tauth$, it thermalizes at the reservoir temperature. 
In other words, after the interaction with the cold/hot reservoir, the system is in the thermal state
\begin{equation}
\rho_{\alpha} = \frac{e^{-\beta_{\alpha} H(\blambda_{\alpha})}}{
\Tr \, e^{-\beta_{\alpha} H(\blambda_{\alpha})}} \;,
\end{equation}
with $\beta_{\alpha}=1/(k_B T_{\alpha})$, 
where $\alpha=c,h$ refers to the cold/hot reservoir, respectively,
and $\blambda_{\alpha}$ is the set of parameters $\blambda$ when the system is in contact with the cold/hot reservoir.
\begin{figure}
    \centering
    \includegraphics[width=0.8\linewidth]{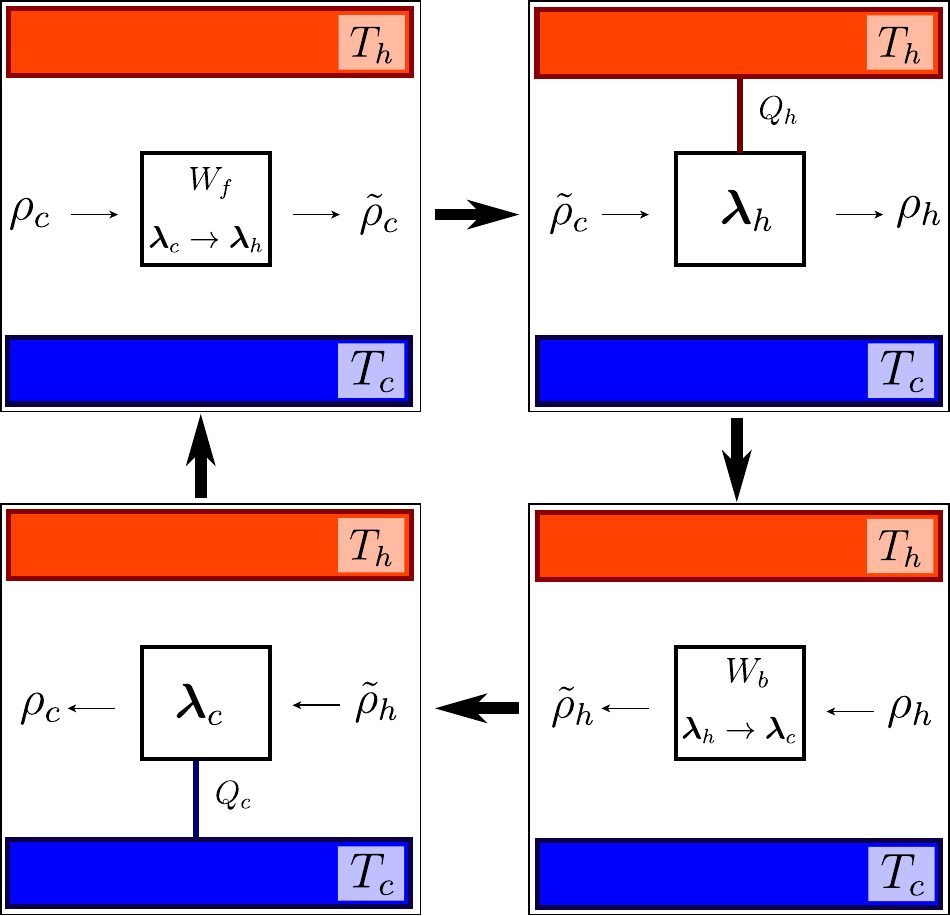}
    \caption{Sketch of the four strokes of a quantum Otto engine.}
    \label{fig:sketch_ciclo}
\end{figure}
 
We implement a four-stroke dynamics on the system that includes two unitary transformations alternated with two thermalization strokes, thus realizing a quantum Otto engine~\cite{quan2007quantum,henrich2007quantum}.
In particular, we prepare the system in the state $\rho_c$ at temperature $T_c$, and we fix $\blambda(0) = \blambda_c$. 
Following that, see sketch in Fig.~\ref{fig:sketch_ciclo}, the four strokes of the cycle are:
\begin{description}
\item[Forward unitary stroke] 
In a time $\tauun$, the external parameters are varied from $\blambda_c$ to $\blambda_h \equiv \blambda(\tauun)$, and the system evolves --- without exchanging heat with the reservoirs --- with a unitary $U_f$, from the initial state $\rho_c$ to the final state $\tilde{\rho}_c = \Uforw \rho_c \Uforw^{\dagger}$. 
\item[Thermalization at $T_h$] 
The system is put in contact with the hot reservoir at temperature $T_h$ for a time $\tauth$, while the Hamiltonian $H_h \equiv H(\blambda_h)$ is fixed (hot dissipative thermalization), reaching the state $\rho_h$.
\item[Backward unitary stroke]  
In a time $\tauun$, the external parameters are changed back from $\blambda_h$ to $\blambda_c$, 
and the system evolves --- without exchanging heat with the reservoirs --- with a unitary $U_b$, from the state $\rho_h$ to the final state $\tilde{\rho}_h = \Uback \rho_h \Uback^{\dagger}$.
\item[Thermalization at $T_c$] 
The system is put in contact with the cold reservoir at temperature $T_c$ for a time $\tauth$, while the Hamiltonian $H_c \equiv H(\blambda_c)$ is fixed (cold dissipative thermalization), returning to the initial state $\rho_c$.
\end{description}
 
During the unitary strokes, only work $W$ is performed and no heat $Q$ is exchanged, while during the thermalization strokes, only heat is exchanged and no work is performed.
The balance between heat and work is governed by
the first law of thermodynamics 
\begin{equation}
   d E = \delta Q - \delta W \;,
   \label{eq:1st_law}
\end{equation}
where $E$ is the internal energy of the system, defined as $E \equiv \braket{H}_\rho$, $\rho$ is the quantum state of the system and $H$ its Hamiltonian. 
Here, we adopt the convention that $\delta Q > 0$ denotes heat absorbed by the system and $\delta W > 0$ denotes work performed by the system. 
According to Eq.~\eqref{eq:1st_law}, we can express the work performed along the forward unitary stroke in terms of Hamiltonian expectation values as
\begin{equation} \label{eq:Wf}
W_f = - \Delta E =  
\big\langle H(\blambda_c)\big\rangle_{\rho_c} - \big\langle H(\blambda_h)\big\rangle_{\tilde{\rho}_c} \;.
\end{equation}
Analogously, we evaluate the work performed along the backward unitary stroke as
\begin{equation}
W_b = - \Delta E =  
\big\langle H(\blambda_h) \big\rangle_{\rho_h} - \big\langle H(\blambda_c) \big\rangle_{\tilde{\rho}_h} \;.
        \label{eq:Wb}
\end{equation}
The total work performed by the system during the cycle is
$W_{\cycle}=W_f+W_b$. For suitable parameters, namely when the energy scales of the two endpoint
Hamiltonians are adequately matched to the reservoir temperatures, this cycle
operates as a heat engine, $W_{\cycle}>0$, as shown for few-level
systems~\cite{Kieu2004,quan2007quantum,henrich2007quantum}, in the nonadiabatic
regime~\cite{Solfanelli_2020}, and for many-body working
media~\cite{piccitto2022ising,Arezzomanybodyengines}. 

It is
therefore meaningful to introduce the power output of a cycle, defined by dividing the total work by the time duration of a cycle, $2\tauun+2\tauth$:
\begin{equation} \label{eq:power_cycle}
\Power_{\cycle} = \frac{W_{\cycle}}{2\tauun+2\tauth} \;,
\end{equation}
where we neglect any possible energetic overheads associated with the external control apparatus. 
We also define the efficiency of the engine as the ratio between the total work of the cycle and the heat $Q_h$ absorbed from the hot reservoir:
\begin{equation} \label{eqn:efficiency}
    \eta = \frac{W_{\cycle}}{Q_h} \;,
\end{equation}
where the heat $Q_h$ can be expressed as
\begin{equation} \label{eq:Qh}
Q_h = 
\braket{H(\blambda_h)}_{\rho_h} - \braket{H(\blambda_h)}_{\tilde{\rho}_c}\;.
\end{equation}

We consider the quantum Otto engine ideal if all the transformations are ideal.
In particular, we require the unitary transformation to be quantum adiabatic (so that the populations associated with the instantaneous energy eigenstates remain unchanged along the adiabatic branches) and the thermalization strokes to be complete (the time $\tauth$ is large enough for the system to reach a thermal state).
 
\section{Free-fermion quantum Otto engine} 
\label{sec:free_fermion_engine}
We consider a lattice of $N$ sites hosting spinless fermions coupled through a generic quadratic Hamiltonian 
\begin{align}
H(\blambda)
=&
\sum_{j,j'=1}^{N}
\mathbf{A}_{j'j}(\blambda)
\left(
\opcdag{j'}\opc{j}
-
\opc{j} \opcdag{j'}
\right)
\nonumber \\
&+
\sum_{j,j'=1}^{N}
\left(
\mathbf{B}_{j'j}(\blambda)
\opcdag{j'}\opcdag{j}
+
\text{H.c.}
\right),
\label{eq:fermionic_hamiltonian}
\end{align}
where $\opc{j}$ and $\opcdag{j}$ are anticommuting fermionic annihilation and creation operators
on the $j$-th site ($j=1,\ldots,N$), 
$\mathbf{A}(\blambda)$ is a Hermitian $N\times N$ matrix and $\mathbf{B}(\blambda)$ a complex antisymmetric $N\times N$ matrix, both depending on the control parameters $\blambda$. 
 
With a Bogoliubov transformation~\cite{mbeng2024quantum,Arezzomanybodyengines} we can always diagonalize the Hamiltonian, for each value of $\blambda$, in terms of $N$ non-interacting Bogoliubov quasiparticle fermions $\opb{\mu}(\blambda)$
\begin{equation}
H(\blambda) = \sum_{\mu=1}^N 2\epsilon_{\mu}(\blambda) \bigg( \opbdag{\mu}(\blambda) \opb{\mu}(\blambda) - \frac{1}{2} \bigg) \;, 
\label{eq:H_bogo}
\end{equation}
where the eigenvalues $\epsilon_{\mu}(\blambda)\ge 0$
depend on $\blambda$. 
Hereafter, we set $\epsilon_{\mu}^\alpha \equiv \epsilon_{\mu}(\blambda_\alpha)$, for $\alpha=c,h$.
 
We can exploit the simple structure of Eq.~\eqref{eq:H_bogo} to evaluate the performed work $W_{\cycle}$ when the two unitary strokes are assumed to be {\em ideal quantum adiabatic} transformations, i.e., they modify the energy eigenvalues without changing the populations of the states.
Indeed, for a quadratic fermionic system~\eqref{eq:fermionic_hamiltonian} in the thermal state, the populations of the $\opb{\mu}$ quasiparticles follow the Fermi--Dirac distribution~\cite{mbeng_quantum_2019}. 
Therefore, the thermal occupation in the starting state $\rho_c$ is:
\begin{equation}
\label{eq:thermalmean}
\Tr \big( \opbdag{\mu}(\blambda_c) \opb{\mu}(\blambda_c) \, \rho_{c} \big) =
\frac{1}{e^{2\beta_c\epsilon_{\mu}^c}+1} \equiv f_F(2\beta_c\epsilon_{\mu}^c) \;,
\end{equation}
where $f_F(x)=1/(e^x+1)$ is the Fermi--Dirac distribution function, and $\beta_c=1/(k_BT_c)$.
The internal energy at the beginning of the transformation reads
\begin{align}
\label{eq:Ec}
E_c \equiv \braket{H(\blambda_c)}_{\rho_c} 
&= \sum_{\mu} 2\epsilon_{\mu}^c \, 
\Big(f_F(2\beta_c\epsilon_{\mu}^c) -\frac{1}{2} \Big) \nonumber \\
&= -\sum_{\mu} \epsilon_{\mu}^c \, 
\tanh(\beta_c\epsilon_{\mu}^c) \;,
\end{align}
where we used that $2(f_F(x)-\frac{1}{2})=-\tanh(x/2)$. 
Assuming an ideal quantum adiabatic process --- whereby the occupation of the states is conserved and only the eigenvalues are changed --- the internal energy after the forward unitary transformation is
\begin{equation}
\tilde{E}_{c} \equiv \braket{H(\blambda_h)}_{\tilde{\rho_c}} = -\sum_{\mu} \epsilon_{\mu}^h \, 
\tanh(\beta_c\epsilon_{\mu}^c) \;.
\label{eq:tEc}
\end{equation}
Hence the work performed in the forward adiabatic unitary stroke is:
\begin{equation}
W_f^\text{id} = \sum_{\mu} (\epsilon_{\mu}^h-\epsilon_{\mu}^c )\, 
\tanh(\beta_c\epsilon_{\mu}^c)\;.
\end{equation}
A similar result can be written for the backward unitary transformation:
\begin{equation}
W_b^\text{id} = \sum_{\mu} 
(\epsilon_{\mu}^c -\epsilon_{\mu}^h)\, 
\tanh(\beta_h\epsilon_{\mu}^h)  \;.
\end{equation}
Collecting all these results we have that
\begin{equation}
W_{\cycle}^\text{id} = W_f^\text{id} + W_b^\text{id} 
= \sum_{\mu} W_{\mu}^\text{id} \;, 
\label{eq:wid}
\end{equation}
where
\begin{equation} \label{eq:wk}
W_{\mu}^\text{id} =
(\epsilon_{\mu}^h - \epsilon_{\mu}^c) 
\big[ \tanh(\beta_c\epsilon_{\mu}^c) -\tanh(\beta_h\epsilon_{\mu}^h) \big] \;.
\end{equation}
 
As for the heat $Q_h$ extracted from the hot reservoir, see Eq.~\eqref{eq:Qh},
if we assume ideal quantum adiabatic unitary transformations, we can substitute the expression for $\braket{H(\blambda_h)}_{\tilde{\rho}_c}$ in Eq.~\eqref{eq:tEc}, finding
\begin{equation}
    Q^{\id}_h = \sum_{\mu}{\epsilon^h_\mu \big[ \tanh(\beta_c\epsilon_{\mu}^c) -\tanh(\beta_h\epsilon_{\mu}^h)  \big]}\;.
\end{equation}
Hence, the ideal efficiency of the engine, see Eq.~\eqref{eqn:efficiency}, is
\begin{equation} \label{eq:idealeff}
\eta^{\id} \!\equiv\! \frac{W^{\id}_{\text{cycle}}}{Q^{\id}_h} = \frac{\sum_{\mu} 
(\epsilon_{\mu}^h - \epsilon_{\mu}^c)\, 
\big[ \tanh(\beta_c\epsilon_{\mu}^c) -\tanh(\beta_h\epsilon_{\mu}^h)   \big]}{\sum_{\mu}{\epsilon^h_\mu\big[\tanh(\beta_c\epsilon_{\mu}^c) -\tanh(\beta_h\epsilon_{\mu}^h) \big]}}\,.
\end{equation}
 
\subsection{Optimal work in ideal Otto cycles} \label{sec:ideal_work}
We now derive the expression for the optimal working point of a free-fermion ideal Otto cycle, by optimizing Eq.~\eqref{eq:wk}.
We observe that, since all the quasiparticle modes are decoupled, the maximum of the ideal work output $W_{\cycle}^\text{id}$ is achieved when each $W_{\mu}^\text{id}$ is independently maximized. 
In turn, $W_{\mu}^\text{id}$ in Eq.~\eqref{eq:wk} is a function of four free parameters: $\epsilon_{\mu}^c, \epsilon_{\mu}^h, T_c, T_h$.
By observing that the hyperbolic tangent, $\tanh(x)$, is a monotonically increasing function of $x$, we conclude that, to perform positive work (see Eq.~\eqref{eq:wk}), we must have:
\begin{equation} \label{eq:work_condition}
\epsilon_{\mu}^h>\epsilon_{\mu}^c>0 
\hspace{5mm} \mbox{and} \hspace{5mm}  
T_c\, \epsilon_{\mu}^h < T_h\, \epsilon_{\mu}^c \;.
\end{equation}

Since the work grows
without bound as $T_h\to\infty$, we fix the temperature of the hot reservoir
to a finite value, $0<T_h<\infty$, and use $k_BT_h$ as the unit of energy. 
The work in Eq.~\eqref{eq:wk} is a decreasing function of $T_c$, and, for
$T_c\to0$, also of $\epsilon_{\mu}^c$. Under the conditions in
Eq.~\eqref{eq:work_condition}, the maximum is therefore achieved for
$T_c\to 0$ and $\epsilon_{\mu}^{c} \to 0$, with $k_B T_c \ll \epsilon_{\mu}^c$
(details in App.~\ref{app:conti_max_work}), obtaining thus
\[
W_{\mu}^\text{id}(T_c\to0,\epsilon^c_\mu\to0) = 2\epsilon^h_\mu f_F(2\beta_h\epsilon^h_\mu) \;.
\]
By maximizing with respect to $\epsilon_{\mu}^{h}$ (still keeping $\epsilon_\mu^c, \,T_c\to 0$) we find that the maximum of the work is achieved when the single-particle eigenvalues are all degenerate (flat bands) and given by
\begin{equation} \label{eqn:zerotemperaturesolution}
\epsilon^{\max,h} = \frac{1}{2}k_B T_h 
\big(1 + W_0(e^{-1}) \big) \approx 0.639\, k_B T_h \;,
\end{equation}
where $W_0(x)$ is the Lambert function, 
leading to a maximum value of the ideal work per mode
\begin{equation} \label{eqn:maxworkideal}
W^{\id,\max}_{\mu} 
= W_0(e^{-1})\,k_BT_h 
\approx 0.279 \,k_B T_h \;,
\end{equation}
which is independent of $\mu$ (see App.~\ref{app:conti_max_work}).
The maximum of the total work is given by
\begin{equation}
W^{\id,\max}_{\cycle} = \sum_{\mu} W^{\id,\max}_{\mu} 
\approx \,0.279 \,N k_B T_h \;,
\end{equation}
thus scaling linearly with the system size and $T_h$. 
 
\begin{figure}[t]
    \centering
    \includegraphics[width=\columnwidth]
    {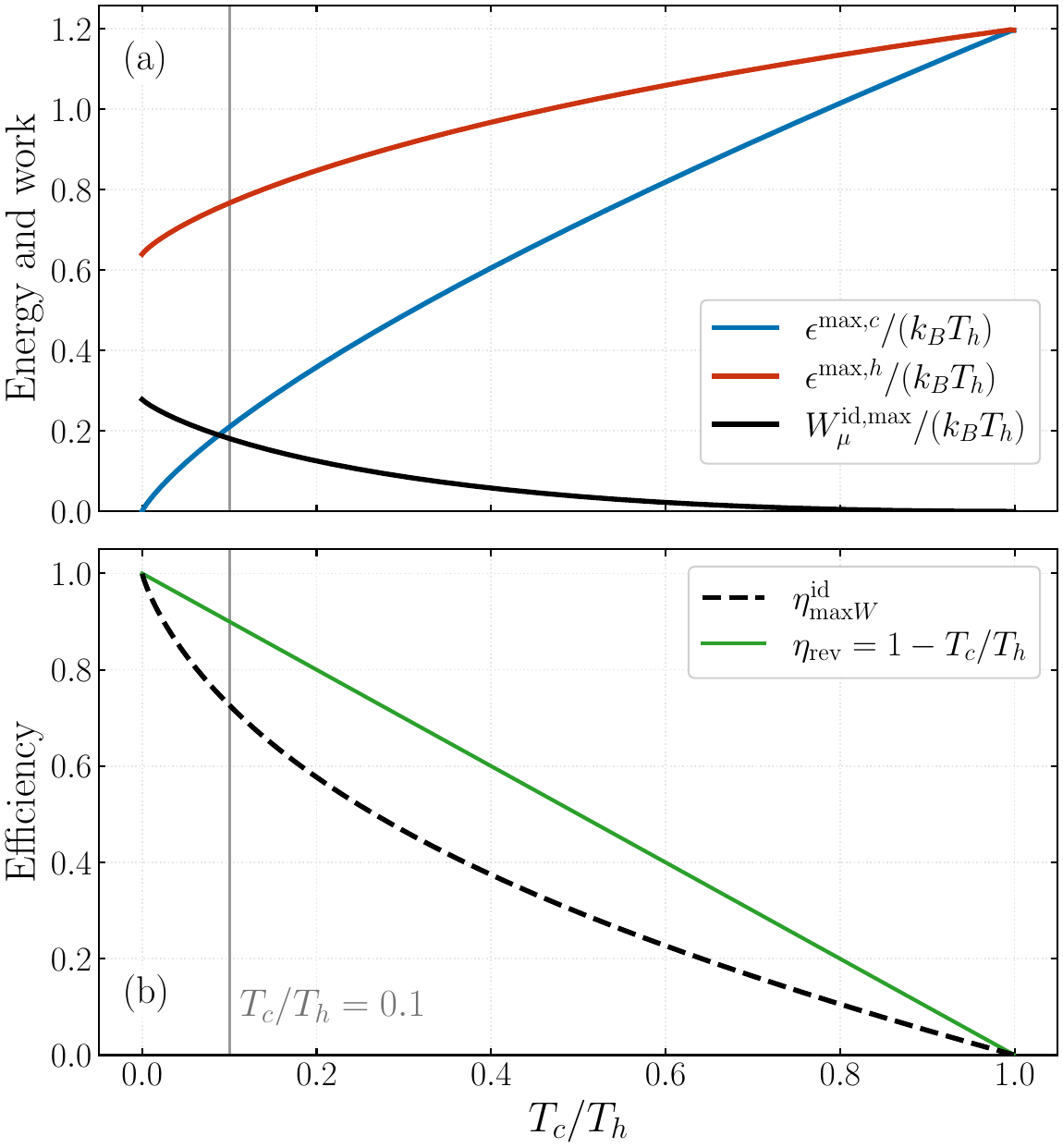}
    \caption{
    Optimal thermodynamic quantities for a single fermionic mode as a
    function of the temperature ratio $T_c/T_h$.
    (a) Optimal energies $\epsilon^{\mathrm{max},c}$ and
    $\epsilon^{\mathrm{max},h}$ and corresponding maximum ideal work
    $W^{\id,\max}_{\mu}$, all expressed in units of $k_B T_h$.
    (b) Efficiency at maximum work $\eta^{\id}_{\mathrm{max}W}$ (dashed
    line), compared with the reversible efficiency
    $\eta_{\mathrm{rev}}=1-T_c/T_h$ (solid line).
    The vertical gray line marks the working point $T_c/T_h=0.1$ used in the rest of the paper, unless otherwise stated.
    }
    \label{fig:ideal_work_vs_temperature_ratio}
\end{figure}

We can extend the above results to the case in which we fix both reservoir temperatures, $0<T_c<T_h$.
In that case, the maximum is reached for two distinct $\mu$-independent values $0<\epsilon^{\max,c}(T_c,T_h)<\epsilon^{\max,h}(T_c,T_h)$ 
--- more details can be found in App.~\ref{app:fixing_T} --- which we
have numerically evaluated, together with the corresponding maximum 
ideal work $W^{\id,\max}_{\mu}(T_c,T_h)$ over a range of temperature ratios $T_c/T_h$.
Assuming we are in this ideal case in which we have two flat bands $\epsilon^h_\mu \equiv \epsilon^{\max,h}$, $\epsilon^c_\mu \equiv \epsilon^{\max,c}$, the expression for the ideal efficiency $\eta^{\id}$ in Eq.~\eqref{eq:idealeff} reduces to
\begin{equation}
\eta^{\id}_{\max W} 
= \frac{W^{\id,\max}_{\cycle}}{Q^{\id}_h} = 
1 -\frac{\epsilon^{\max,c}}{\epsilon^{\max,h}}\;. 
\end{equation}
In the flat-band case, the Otto efficiency $1-\epsilon^c/\epsilon^h$ reaches the
reversible Carnot value $\eta_{\mathrm{rev}}=1-T_c/T_h$ only at the boundary of
Eq.~\eqref{eq:work_condition}, $T_c\,\epsilon^h=T_h\,\epsilon^c$, where the
thermal populations at the two endpoints coincide and the work vanishes.
Maximizing the work therefore necessarily yields
$\eta^{\id}_{\max W}<\eta_{\mathrm{rev}}$.
In Fig.~\ref{fig:ideal_work_vs_temperature_ratio}, we show the optimal energies
$\epsilon^{\mathrm{max},c}$ and $\epsilon^{\mathrm{max},h}$, together with the
corresponding maximum ideal work $W^{\id,\max}_{\mu}$, as functions of
$T_c/T_h$, with all energies expressed in units of $k_B T_h$. We also show the
efficiency at maximum work, $\eta^{\id}_{\mathrm{max}W}$, and compare it with
$\eta_{\mathrm{rev}}$.
\section{Transverse-field Ising engine}
\label{sec:ising_engine}
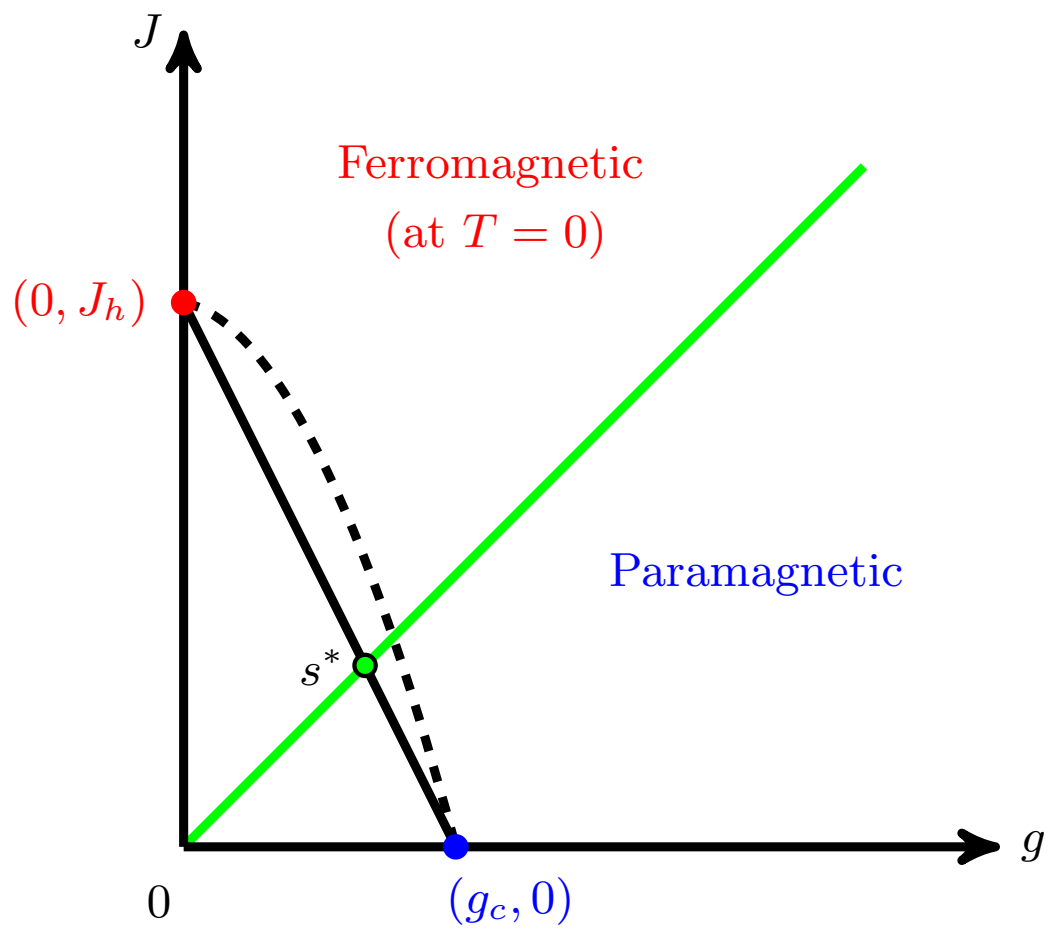
\begin{figure}[!ht]
\centering
\begin{tikzpicture}[scale=1.8, every node/.style={scale=1.0}, square/.style={regular polygon, regular polygon sides=4}]
\draw [line width = 0.6mm, draw=green] (0,0) -- (2.5,2.5) node[right, black] {};
\draw [line width = 0.6mm, draw=black] (0,2) -- (1,0) node[right, black] {};
\draw [line width = 0.6mm, draw=black, ->,>=stealth'] (0,0) -- (3,0) node[right, black] {$g$};
\draw [line width = 0.6mm, draw=black, ->,>=stealth'] (0,0) -- (0,3) node[left, black] {$J$};
\draw[black,line width = 0.6mm, dashed] (0.,2.) .. controls (0.25,1.95) and (0.6,1.5) .. (1.,0.) ;
\node[circle,draw=blue,thick,fill=blue,inner sep=0pt,minimum size=4pt,label=below:{}] (a) at (1,0) {};
\draw (0.9,-0.2) node[right, text=blue]{$(g_c,0)$};
\node[circle,draw=red,thick,fill=red,inner sep=0pt,minimum size=4pt,label=below:{}] (a) at (0,2) {};
\draw (-0.7,2) node[right, text=red]{$(0,J_h)$};
\draw (0.36,0.66) node[right, text=black]{$s^*$};
\node[circle,draw=black,thick,fill=green,inner sep=0pt,minimum size=4pt,label=below:{}] (a) at (2/3,2/3) {};
\draw (-0.2,-0.2) node[right, text=]{$0$}; 
\draw (0.5,2.5) node[right, text=red]{Ferromagnetic}; 
\draw (0.67,2.25) node[right, text=red]{(at $T=0$)}; 
\draw (1.5,1.0) node[right, text=blue]{Paramagnetic}; 
\end{tikzpicture}
\caption{The ground-state phase diagram of the quantum Ising chain in Eq.~\eqref{eq:isinghamiltonian}. 
The green solid line $J=g$ denotes the quantum critical line. 
The lines connecting the point $(g_c,0)$ in the quantum paramagnetic phase to the point $(0,J_h)$ in the ferromagnetic phase indicate two possible curves $\blambda(t)$ associated with a forward unitary transformation. 
The solid straight line is the path used throughout this work.
The critical value of the interpolation parameter $s^*=g_c/(g_c+J_h)$ is shown.}
\label{fig:phase_diagram}
\end{figure}

We now consider a translationally invariant quantum Ising spin chain, whose Jordan--Wigner fermionic representation is a particular case of the general quadratic fermionic model in Eq.~\eqref{eq:fermionic_hamiltonian}. 
The Hamiltonian reads~\cite{mbeng2024quantum}
\begin{equation} \label{eq:isinghamiltonian}
H(g,J) = \!\sum_j\! \Big[ -J 
\big( \opcdag{j} \opc{j+1} + \opcdag{j} \opcdag{j+1} + \text{H.c.} \big) +g \big( \opcdag{j} \opc{j} - \opc{j} \opcdag{j} \big) \Big]\,,
\end{equation}
with $J,g \geq 0$.
Differently from previous works~\cite{piccitto2022ising, Arezzomanybodyengines}, we account for the possibility of separately varying the coupling $J$ and the transverse field $g$.
In practice, we imagine the unitary transformation as realized by a two-dimensional curve $\blambda(t)$ connecting two points in the $(g,J)$ phase diagram of the Ising model, as sketched in Fig.~\ref{fig:phase_diagram}.
 
We can analytically diagonalize the Hamiltonian by a Bogoliubov transformation, 
leading to momentum-space quasiparticles with energies \cite{mbeng2024quantum}
\begin{equation} \label{eqn:epsilon_k}
\epsilon_k(g, J) = \sqrt{g^2 + J^2 - 2gJ\cos k} \;.
\end{equation}
Here $k =\pm (2n-1)\pi/N$ with $n=1\cdots N/2$ in the even fermion-parity sector, and $k =0, \pi, \pm 2n\pi/N$ with $n=1,\ldots, (N/2-1)$ in the odd fermion-parity sector~\cite{mbeng2024quantum}. 
 
The quantum Ising chain exhibits a critical point at $g = J$, separating the quantum paramagnetic phase ($g>J$) from the ferromagnetic phase ($g<J$), see Fig.~\ref{fig:phase_diagram}: this criticality is accompanied by a closure of the gap in the energy spectrum for $k\to 0$.
 
Let us imagine a forward unitary transformation that brings the system, in a total time $\tauun$, from the initial point $(g_c, J_c)=\blambda(0)$ in the phase diagram to the final point $(g_h, J_h)=\blambda(\tauun)$. 
The backward stroke traverses the same path in the opposite direction.
For simplicity of notation, we set $\epsilon_k(g_\alpha, J_\alpha) \equiv \epsilon_k^\alpha$ for $\alpha = c, h$. At finite size, thermal averages involve both the even and odd fermion-parity sectors of the Hilbert space, which have different sets of wave vectors~\cite{mbeng2024quantum}.
We restrict the unitary dynamics to the even sector, with $k=\pm(2n-1)\pi/N$,
$n=1,\ldots,N/2$.
The thermal factors for the Gibbs state projected onto this sector are derived in App.~\ref{app:thermoising}. In the numerical analysis below, we use their thermodynamic-limit expressions. At fixed temperatures and fixed endpoint energy scales, the resulting finite-size corrections to these thermal factors
vanish exponentially with $N$.
Following Eq.~\eqref{eq:wk}, the ideal work then reads
\begin{equation} \label{eq:idealworkgeneral}
W_{\cycle}^{\id} = \sum_{k}
\left(\epsilon_k^h-\epsilon_k^c\right)
\left[\tanh(\beta_c\epsilon_k^c) -\tanh(\beta_h\epsilon_k^h) \right] \;,
\end{equation}

As we described in Sec.~\ref{sec:ideal_work}, the maximum work is attained for $\epsilon^h_k \equiv\epsilon^{\max,h}(T_c/T_h)$, and $\epsilon^c_k\equiv  \epsilon^{\max,c}(T_c/T_h)$, identical for all wave vectors $k$,
with $\epsilon^{\max,h}>\epsilon^{\max,c}$.
By looking at Eq.~\eqref{eqn:epsilon_k}, there are two possible routes to obtain a $k$-independent $\epsilon^\alpha_k$:
\begin{enumerate}
\item One route is to consider $J_c = J_h = 0$ so that $\epsilon^{c}_k = g_c$, and $\epsilon^{h}_k = g_h$, or, vice-versa, $g_c = g_h = 0$ with $\epsilon^{c}_k = J_c$, and $\epsilon^{h}_k = J_h$: in both cases, the Hamiltonian commutes with itself for different $t$, hence the dynamics is trivial (and, incidentally, the path $\blambda(t)$ does not cross the critical line $J=g$). 
\item
A more interesting route is attained when $J_c = g_h =0$ so that $\epsilon^c_k = g_c$ and $\epsilon^h_k = J_h$, 
and the evolution connects two flat-band situations. 
We can consider, for example, a forward evolution from $(g_c,0) \to (0,J_h)$, with $J_h \equiv \epsilon^{\max,h}(T_c/T_h)$ and $g_c \equiv \epsilon^{\max,c}(T_c/T_h)$ according to the values reported in Fig.~\ref{fig:ideal_work_vs_temperature_ratio}.
This case is interesting because the evolution will cross the critical line $J=g$. 
\end{enumerate}  
Using the optimal couplings $J_h$ and $g_c$ we can write the ideal work as
\begin{equation} \label{eqn:idealworkottoengine}
W^{\id,\max}_{\cycle}
= N (J_h-g_c)
\big[ \tanh(\beta_c g_c) - \tanh(\beta_h J_h)\big] \;.
\end{equation}
In the following, we restrict ourselves to the straight path connecting the two fixed endpoints, parametrized as
\begin{equation}
\blambda(t) =
\big[ g_c (1-s(t)), J_h s(t) \big]
\end{equation}
with
\begin{equation}
    s(0)=0 \;,\quad s(\tau_{\rm un})=1 \;,
\end{equation}
and optimize only the temporal parametrization $s(t)$ along the path. By restricting the evolution to this path, we
isolate the effect of optimizing the temporal schedule from that of changing
the overall Hamiltonian energy scale. Indeed, a uniform increase of the
energy scales would trivially shorten the characteristic dynamical times by
increasing the relevant gaps, without representing a genuine improvement of
the control protocol.
  
\section{Optimization of the cycle work} \label{subsec:ising_work_real}
In this section, we present the quantum annealing (QA) protocol we adopted to numerically optimize the performance of the engine and the resulting cycle work. 
Because of the closing spectral gap at the quantum critical point, the choice of the protocol is crucial to avoid nonadiabatic excitations in the system. 
 
\subsection{The optimization protocol}
In continuous-time QA we consider a time-dependent interpolating Hamiltonian
\begin{equation}
    H(t) = s(t) H_{\text{targ}} + ( 1 - s(t)) H_{\text{mix}}\;,
\end{equation}
where $s(0)=0$, $s(\tau_{\rm un})=1$ and $\tau_{\rm un}$ is the total annealing time.  We start with the ground state $\ket{\psi(0)} \equiv \ket{\psi_0}$ of $H_{\text{mix}}$ and we perform the evolution according to the Schr\"odinger equation
\begin{equation} \label{eqn:SE}
    i\hbar \frac{d }{d t} |\psi(t)\rangle = H(t) |\psi(t)\rangle \;.
\end{equation}
The aim is to optimize $s(t) \in [0,1]$ in such a way that the final energy,
\begin{equation}
    E^{\text{fin}}(\tau_{\rm un}) = \bra{\psi(\tau_{\rm un})}H_\text{targ}\ket{\psi(\tau_{\rm un})}\;,
\end{equation}
is minimized. 
We can apply this formalism to our Ising Hamiltonian in Eq.~\eqref{eq:isinghamiltonian}. 
 
\textbf{Forward unitary stroke.} 
First we focus on the forward unitary stroke from $(g_c,0) \to (0,J_h)$; therefore, we define
\begin{equation} \label{eqn:hamiltoniansdefinition_forward}
H_\text{mix} \equiv H_g\;, \qquad
H_\text{targ} \equiv H_J \;,
\end{equation}
with 
\begin{subequations}\label{eq:HgHj}
\begin{eqnarray} \label{eqn:hamiltoniansdefinition2}
H_g & = & g_c\sum_{j=1}^{N} \big( \opcdag{j} \opc{j} - \opc{j} \opcdag{j} \big)\;,\\
H_J & = & - J_h\sum_{j=1}^{N}  
\big( \opcdag{j} \opc{j+1} + \opcdag{j} \opcdag{j+1} + \text{H.c.} \big)\; ,
\end{eqnarray}
\end{subequations}
obtaining a forward annealing Hamiltonian,
\begin{equation}
H_f(t) = s(t)\, H_J + \big(1 - s(t)\big) \, H_g \;.
\end{equation}
 
Our goal is to find an {\em optimal} choice for the schedule $s(t)$ which improves the quality of the final state obtained in a finite time $\tauun$, with respect to the obvious linear choice $s^{\mathrm{lin}}(t)=t/\tauun$. 
Postponing for a moment the technicalities involved in this {\em functional optimization} problem, we point out an important difference with respect to the standard QA where the focus is on the unitary dynamics of {\em pure states}: here, we want to deal with the unitary dynamics of {\em thermal states}, which poses an immediate extra difficulty. 
 
Suppose that we can construct the unitary operator $U_f(\tauun)$ associated with the chosen schedule $s(t)$ in a given finite time $\tauun$; the work performed along the forward unitary stroke, see Eq.~\eqref{eq:Wf}, should be evaluated as
\begin{equation} \label{eq:w_fd}
W_f = \Tr\left(\rho_c H_g\right) -
    \Tr\big( U_f \rho_c U_f^\dagger H_J\big) \;.
\end{equation}
The standard pure-state QA would require calculating this quantity when $\rho_c=|\psi_g\rangle\langle \psi_g|$, where $|\psi_g\rangle$ is the ground state of $H_g$, while we want to calculate it for the thermal state $\rho_c$, which we take in the even fermion-parity sector, $\rho_c=P_+e^{-\beta_c H_g}/Z_{c,+}$, with $P_+$ the projector on that sector (see App.~\ref{app:thermoising}). 
The final result --- derived in detail in App.~\ref{app:thermoising} --- is that we can account for all finite-temperature statistical weights by writing
\begin{align} \label{eqn:W_f_thermal}
W_f &= \big(-g_c N - \langle \psi_g | U_f^\dagger H_J U_f |\psi_g\rangle \big) \, f_N(\beta_c g_c) \nonumber \\
&= N\big[ (J_h-g_c) - \epsilon^f_{\res}(\tauun) \big] \, f_N(\beta_c g_c) \;,
\end{align}
where 
\begin{equation}
\epsilon^f_{\res}(\tauun) = \frac{1}{N} \Big[
\langle \psi_g | U_f^\dagger(\tauun) H_J U_f(\tauun) |\psi_g\rangle-E_0^J \Big] \;,
\end{equation} 
is the standard QA residual energy per site over the target ground-state energy $E_0^J=-J_h N$, and 
\begin{equation}
\label{eq:sizethermalfactor}
f_N(x) = \frac{\tanh x+\tanh^{N-1}x}{1+\tanh^N x} \xrightarrow{N\to \infty} \tanh(x) \;,
\end{equation}
accounts for the statistical Boltzmann factors; the pure-state evolution results are obtained by setting $f_N\equiv 1$ in Eq.~\eqref{eqn:W_f_thermal}. Since $f_N(\beta_c g_c)$ does not depend on the protocol, the schedule that
minimizes the pure-state residual energy $\epsilon^f_{\res}$ also maximizes the
work extracted from the thermal state $\rho_c$.
 
\textbf{Backward unitary stroke.} 
Similar steps can be followed for the backward unitary stroke, where we set
\begin{equation} \label{eqn:hamiltoniansdefinition_back}
H_\text{mix} \equiv H_J\;, \qquad
H_\text{targ} \equiv H_g \;,
\end{equation}
obtaining a backward annealing Hamiltonian
\begin{equation}
H_b(t) = \big[1 - s_b(t)\big] \, H_J + s_b(t) \, H_g\;, 
\end{equation}
where the schedule, once again, satisfies $s_b(0)=0$, and $s_b(\tauun) = 1$. 
If $U_b(\tauun)$ is the unitary evolution operator, the backward stroke work can be
written as
\begin{align} \label{eq:w_bwd}
W_b &= \Tr\left(\rho_h H_J\right) -
    \Tr\big( U_b \rho_h U_b^\dagger H_g\big) \nonumber \\
    &= 
    \big(-J_hN  - \langle \psi_J | U_b^\dagger H_g U_b |\psi_J\rangle \big) \, f_N(\beta_h J_h) \nonumber \\
&= N\big[ (g_c-J_h) - \epsilon^b_{\res}(\tauun) \big] \, f_N(\beta_h J_h) \;,
\end{align}
where $|\psi_J\rangle$ is the ground state of $H_J$ and
\begin{equation}
\epsilon^b_{\res}(\tauun) = \frac{1}{N}\Big[ 
\langle \psi_J | U_b^\dagger(\tauun) H_g U_b(\tauun) |\psi_J\rangle-E_0^g \Big] \;,
\end{equation} 
is the QA residual energy per site over the target ground-state energy $E_0^g=-g_c N$.
 
\textbf{The cycle work at finite-$\tauun$.}
Grouping these two results, and using the thermodynamic-limit approximation $f_N(x)\approx\tanh(x)$ for the thermal factors, while retaining the finite-size unitary dynamics, we can write the total work per site 
$w_{\cycle}=(W_f+W_b)/N$ produced in the cycle as
\begin{equation}
w_{\cycle} = w^{\id,\max}_{\cycle}
-\Delta_{\mathrm{n.a.}},
\end{equation}
where we recognize the first contribution to be the ideal maximum work per site produced in the cycle $W^{\id,\max}_{\cycle}/N$ [see Eq.~\eqref{eqn:idealworkottoengine}],
\begin{equation}
w^{\id,\max}_{\cycle} = (J_h-g_c) \big[\tanh(\beta_c g_c) - \tanh(\beta_h J_h)\big] \;,
\end{equation}
and a nonadiabatic loss term $\Delta_{\mathrm{n.a.}}$, which accounts for the detrimental effect of nonadiabatic finite-$\tauun$ internal excitations captured by the two residual energies:
\begin{equation}
\Delta_{\mathrm{n.a.}} = 
    \epsilon_{\res}^f(\tauun) \tanh(\beta_c g_c) + 
    \epsilon_{\res}^b(\tauun) \tanh(\beta_h J_h) .
\end{equation}
In the following, we drop the subscript and write
$w\equiv w_{\cycle}$ and $w^{\id,\max}\equiv w^{\id,\max}_{\cycle}$.

\textbf{The schedule optimization problem.} 
We now need to address the schedule optimization problem, i.e., how to devise an $s(t)$ and an $s_b(t)$ such that $\epsilon_{\res}^f(\tauun)$ and $\epsilon_{\res}^b(\tauun)$
are both minimized. 
Luckily, the Ising-model duality between $H_J/J_h$ and $H_g/g_c$ makes the second optimization unnecessary, see App.~\ref{app:thermoising} for details: if an optimal $s^{\opt}(t)$ is known for the forward protocol, then
\begin{equation}
    s_b^{\opt}(t) = 1 - s^{\opt}(\tauun-t) \;,
\end{equation}
is the optimal backward schedule, and the corresponding residual energies are such that:
\begin{equation}
    \frac{1}{J_h} \epsilon_{\res}^f(\tauun) =  \frac{1}{g_c} \epsilon_{\res}^b(\tauun) \;.
\end{equation}
Since the two residual energies are measured on target Hamiltonians with different energy scales, we normalize each of them by its own scale and define the dimensionless residual energy per site
\begin{equation} \label{eqn:e_res_def}
    \eres(\tauun) = \frac{1}{J_h} \epsilon_{\res}^f(\tauun) =  \frac{1}{g_c} \epsilon_{\res}^b(\tauun) \;,
\end{equation}
and re-express the nonadiabatic loss term as
\begin{equation} \label{eqn:w_cycle_fin}
\Delta_{\mathrm{n.a.}} = \big[ J_h  \tanh(\beta_c g_c) +g_c \tanh(\beta_h J_h) \big] \, \eres(\tauun) \;.
\end{equation}
 
\textbf{Digitization of the schedule.}
We are finally led to the crucial problem of optimizing the schedule $s(t)$ which will allow us to minimize the nonadiabatic contributions present in $\eres(\tauun)$. To do that, we use the standard route of simplifying the functional optimization into a finite-dimensional minimization, through digitization and Trotterization~\cite{Nature_dQA,Mbeng_dQA_PRB2019,mbeng_quantum_2019,pecci2024beyond}, leading us to a QAOA~\cite{Farhi_arXiv2014} {\em Ansatz} for the evolution operator $U_f$.
 
The idea is to divide the dynamics into a large number $P$ of small evolution steps of length $\Delta_t = \tau_{\rm un} / P$. 
For our simulations we choose $\Delta_t = 0.1 \hbar/(k_B T_h)$ as in Ref.~\cite[App.~B]{arezzo2026}, with a Trotter error --- controlling how closely the digitized protocol reproduces a continuous-time schedule $s(t)$ --- which we find to be negligible for the main results of our work.
We define the discrete times $t_p = \left(p - \frac{1}{2}\right) \frac{\tau_{\rm un}}{P}$ and write the discretized evolution operator as 
\begin{equation}
    U_f(\tau_{\rm un}) = \prod_{p=1}^P{e^{-\frac{i}{\hbar}H(s(t_p))\Delta_t}}\;.
\end{equation}
Defining
\begin{equation} \label{eqn:simpleangles}
    \theta^J_p = s(t_p) \, \Delta_t\;,\;\; \theta^g_p = (1 - s(t_p))\, \Delta_t\;,
\end{equation}
we Trotter-decompose the evolution operator, finding
\begin{equation} \label{eq:trotterized_control_protocol}
U_f(\btheta) = \prod_{p=1}^{P}
    e^{-\frac{i}{\hbar} H_g\,\theta^g_p}
    e^{-\frac{i}{\hbar} H_J\, \theta^J_p} \;,
\end{equation}
with $\btheta=(\theta^g_1,\ldots,\theta^g_P,\theta^J_1,\ldots,\theta^J_P)$. 

\textbf{The CRAB {\em Ansatz}.} 
We now make an {\em Ansatz} for the protocol $s(t)$ 
according to the Chopped RAndom Basis (CRAB) approach of Ref.~\cite{Caneva_PRA2011} and recently used in Ref.~\cite{arezzo2026}. 
We define $s(t)$ as follows:
\begin{equation} \label{eqn:schedule_CRAB}
s(t) = \frac{t}{\tau_{\rm un}} + \sin\bigg({\frac{\pi \, t}{\tau_{\rm un}}}\bigg) \sum_{n=1}^{N_c} \Big[ {C^s_n \sin(\omega_n t)} + 
C^c_n 
\cos(\omega_n t) \Big] \;.
\end{equation}
In essence, we add a finite number, $2N_c$, of Fourier components
with random frequencies $\omega_n$ to the linear schedule $s^{\mathrm{lin}}(t)=t/\tauun$. 
The goal is to optimize the $2N_c$ variational parameters $\C=(\C^s,\C^c)$, with $\C^{s/c}=(C^{s/c}_1,\ldots,C^{s/c}_{N_c})$ --- entering into the QAOA angles $\btheta(\C)$, see Eqs.~\eqref{eqn:simpleangles}-\eqref{eqn:schedule_CRAB} --- by minimizing the residual energy as defined in Eq.~\eqref{eqn:e_res_def},
\begin{equation} \label{eq:Cresen}
\eres(\C)= \frac{1}{J_hN}
\Big( \bra{\psi_g} U_f^\dagger(\C)\, H_J \, U_f(\C)
\ket{\psi_g} - E_0^J \Big) \,.
\end{equation}
Since the {\em Ansatz} in Eq.~\eqref{eqn:schedule_CRAB} does not by itself
guarantee $0\le s(t)\le1$, i.e., nonnegative angles $\theta^J_p$ and
$\theta^g_p$ in Eq.~\eqref{eqn:simpleangles}, we actually minimize the penalized cost
\begin{equation} \label{eq:penalty}
\left\{
\begin{aligned}
\mathcal{E}(\C) &= J_hN\,\eres(\C) + \mathcal{E}_{\rm pen}(\C) \\
\mathcal{E}_{\rm pen}(\C) &= \frac{\lambda}{\Delta_t^2}\sum_{p=1}^{P}
\Big( \big[\theta^J_p\big]_-^2 + \big[\theta^g_p\big]_-^2 \Big)
\end{aligned}
\right. \;,
\end{equation}
where $[x]_-\equiv\max(0,-x)$ and $\lambda=10^8\,k_BT_h$. 
The penalty term vanishes for nonnegative angles
and, being quadratic, keeps the cost differentiable, so that its gradient
is added analytically to that of the residual energy. The optimized schedules satisfy the bounds up to violations smaller than
$3\times10^{-5}$.
For the numerical simulations, we follow the procedure described in
Ref.~\cite{arezzo2026}. We take $N_c=20$ and draw the dimensionless
frequencies $x=\omega_n\tau_{\rm un}/\pi$ from the Gamma distribution
\begin{equation}
\Gamma(x\geq0;a,b)
=
\frac{x^{a-1}e^{-x/b}}
{b^{a}\,\Gamma(a)} \;,
\end{equation}
where $a>0$ controls the behavior at small $x$, while $b>0$
sets the characteristic scale at large $x$. For $a>1$, the distribution is peaked
at $x_{\max}=(a-1)b$. Throughout our numerical analysis, we used
$a=3/2$ and $b=4$, corresponding to $x_{\max}=2$.
 
The outcome of a single CRAB optimization depends on the particular
realization of the random frequencies $\omega_n$. To reduce this
dependence, for each set of physical parameters and each unitary-stroke
duration $\tauun$ we perform $N_{\rm real}=20$ independent optimizations,
each with its own draw of the frequencies, and retain the one yielding
the lowest optimized residual energy. 
We refer to this best-of-$N_{\rm real}$ procedure as a single 
optimization \emph{run}, and denote its outcome by
\begin{equation}
e_{\res,r}
=
\min_{i=1,\ldots,N_{\rm real}} e_{\res,r,i} \;,
\end{equation}
where $i=1,\ldots, N_{\rm real}$ labels the frequency realization and $r=1,\ldots, N_{\rm run}$ the run.
To assess the reproducibility of the optimization, we repeat the whole
procedure $N_{\rm run}=10$ times with independent random seeds, obtaining
$N_{\rm run}$ statistically independent values $e_{\res,r}$.
Since the residual energy spans several orders of magnitude, its natural
scale is logarithmic. We therefore characterize the typical outcome by the
geometric mean, i.e., the exponential of the average of
$\ln e_{\res,r}$,
\begin{equation} \label{eq:eres_logmean}
\overline{e}_{\res} =
\exp\bigg( \frac{1}{N_{\rm run}} \sum_{r=1}^{N_{\rm run}}
\ln e_{\res,r} \bigg) \;,
\end{equation}
and the spread of the outcomes by the full range spanned by the runs,
\begin{equation} \label{eq:eres_range}
\eres^{\min} = \min_{r} e_{\res,r} \;, \qquad
\eres^{\max} = \max_{r} e_{\res,r} \;.
\end{equation}
The geometric mean describes the typical order of magnitude of the
residual energy and, unlike the arithmetic mean, is not dominated by the
worst runs. For the spread we do not use a band of one standard deviation:
close to the controllability threshold, the distribution of
$\ln e_{\res,r}$ over independent runs is strongly skewed, and a
$\pm\sigma$ band, either linear or logarithmic, can extend beyond the
values actually attained by the runs. The min--max range, instead, always
corresponds to actual optimization outcomes and contains the geometric
mean by construction.
 
Since the cycle work per site depends linearly on the residual energy,
see Eq.~\eqref{eqn:w_cycle_fin},
\begin{equation} \label{eq:A_def}
\begin{aligned}
w &= w^{\id,\max} - A\, \eres \\
A &= J_h \tanh(\beta_c g_c) + g_c \tanh(\beta_h J_h)  \;,
\end{aligned} 
\end{equation}
the work shortfall $w^{\id,\max}-w \equiv \Delta_{\mathrm{n.a.}} = A \, \eres$ due to nonadiabatic excitations is proportional to the residual energy. 
We therefore apply the same statistics to it: the
typical work is obtained from the geometric mean of the shortfall, and
since the relation is monotonically decreasing, the extremes of the
residual energy map onto the opposite extremes of the work,
\begin{equation}
\begin{aligned} \label{eq:work_logmean_range}
\overline{w} &= w^{\id,\max} - A\,\overline{e}_{\res} \;\\
w &\in \big[\, w^{\id,\max} - A \, \eres^{\max},\;
w^{\id,\max} - A \, \eres^{\min} \,\big] \;.
\end{aligned}
\end{equation}
The same holds for the cycle power $\Power_{\cycle}$, which at fixed $\tauun$ is obtained by
dividing the work by the cycle duration $2(\tauun+\tauth)$, see Eq.~\eqref{eq:power_cycle}.
 
The time discretization and Trotterization of the dynamics are very useful for the finite-dimensional optimization problem. As shown in App.~\ref{app:optimization_evolution}, the residual energy decomposes into a sum of $N/2$ independent contributions, each involving only products of $2\times2$ matrices, Eq.~\eqref{eq:forward_residual_fidelities}. 
This gives easy access to the analytical gradients of $\eres(\btheta(\C))$~\cite[App.~C and D]{wang2025exponential}, which are passed to the Broyden--Fletcher--Goldfarb--Shanno (BFGS) algorithm~\cite{nocedal1999numerical} used to minimize the residual energy.  
 
\subsection{Maximum work output results}
We now compare the results obtained with an optimized schedule $s^{\rm opt}(t)$, see 
Eq.~\eqref{eqn:schedule_CRAB}, with those obtained with the linear schedule
$s^{\lin}(t)=t/\tauun$. 
In the following we take $k_BT_h$ as the energy
unit, $T_c=0.1\,T_h$, and, 
correspondingly (see Fig.~\ref{fig:ideal_work_vs_temperature_ratio}),   the work-optimal energy scales
$J_h\approx0.767\,k_BT_h$ and $g_c\approx 0.210\,k_BT_h$, for which the critical point is crossed at $s^*=g_c/(g_c+J_h)\approx 0.215$. 
Times are measured in units of $\hbar/(k_BT_h)$ and powers in units of $(k_BT_h)^2/\hbar$.
 
\begin{figure}[!t]
    \centering
    \includegraphics[width=\columnwidth]{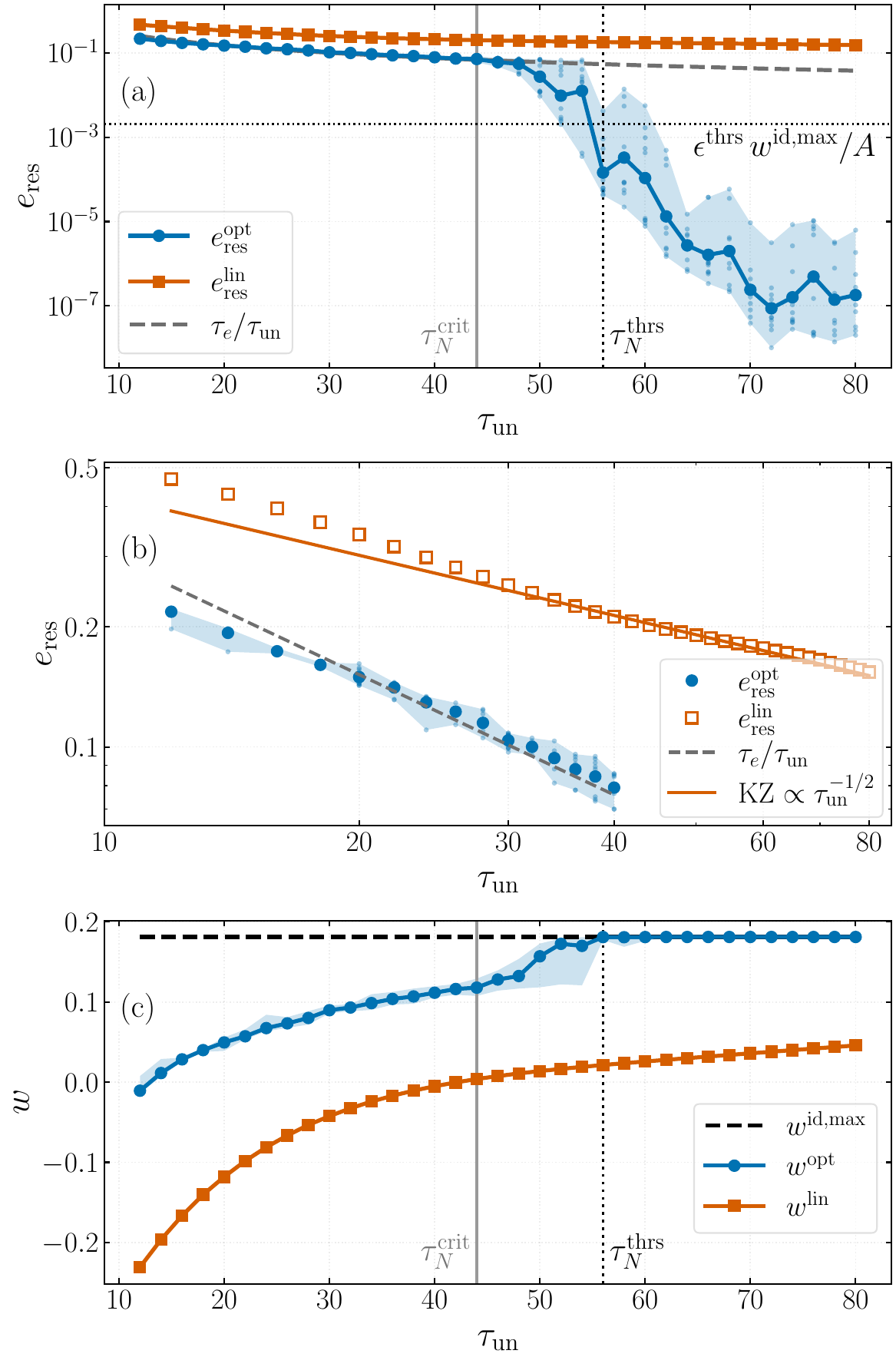}
    \caption{ 
    (a) Forward residual energy per site $\eres$ as a function of the
    unitary-stroke duration $\tauun$ for $N=30$. 
    Circles and squares correspond to the
    optimized ($\eres^{\opt}$) and linear ($\eres^{\lin}$) protocols. 
    For $\eres^{\opt}$, circles
    denote the geometric mean $\overline{e}_{\res}$ in Eq.~\eqref{eq:eres_logmean} for $N_{\rm run}=10$ runs, each
    with the best of $N_{\rm real}=20$ CRAB frequency realizations;
    the shaded band shows the full range of the runs, Eq.~\eqref{eq:eres_range}, and small dots
    the individual runs. 
    The gray dashed line is Eq.~\eqref{eqn:eres_opt}.
    The dotted horizontal line is the threshold value 
    $\eres=\threshold\,w^{\id,\max}/A$  
    with $\threshold=10^{-2}$, corresponding
    to a $1\%$ loss of cycle work. 
    (b) The power-law regime on a logarithmic scale: $\eres^{\opt}(\tauun\le 40)$ (gray dashed line is Eq.~\eqref{eqn:eres_opt}), and  $\eres^{\lin}$ (solid line is Eq.~\eqref{eq:kz_linear}). 
    (c) Cycle work per site; the black dashed line denotes the ideal value
    $w^{\id,\max}\approx0.181\,k_BT_h$, see
    Fig.~\ref{fig:ideal_work_vs_temperature_ratio}. In (a) and (c), the vertical solid gray lines mark $\tau^{\crit}_N\approx 44$, where $\eres^{\opt}$ departs from Eq.~\eqref{eqn:eres_opt}, and the vertical dotted lines mark the numerical threshold time $\tau^{\thrs}_{N=30}=55$, see Fig.~\ref{fig:taucrit_numeric}, beyond which $\overline{e}_{\res}$ stays below the threshold. 
    Times are in units of $\hbar/(k_BT_h)$.}
    \label{fig:cycle_work_comparison}
\end{figure}
 
Figure~\ref{fig:cycle_work_comparison} summarizes the results for $N=30$.
The residual energy $\eres^{\opt}$ of the optimized protocol, panel~(a), first decreases as $1/\tauun$,  
and then departs, with a rapid drop, from this behavior around the
finite-size critical time $\tau^{\crit}_{N=30}\approx 44$,
Eq.~\eqref{eq:tau_crit}, which lower-bounds the control time required for
full controllability of the system~\cite{Arezzodigital2025}, i.e., for
$\eres^{\opt}=0$~\footnote{With bang-bang controls~\cite{Brady_PRL2021}, exact
ground-state preparation is possible at a specific evolution time
$\tauun(N)\gtrsim\tau^{\crit}_N$~\cite{Arezzo_inpreparation}.}.
The rapid drop of $\eres^{\opt}$ over many decades allows us to define a
{\em threshold} time $\tau^{\thrs}_N$ beyond which the work differs from the
ideal one by less than a small fraction $\threshold$, which we take to be
$10^{-2}$, see Sec.~\ref{sec:numerical_critical_time} for more details. 
 
The linear schedule residual energy $\eres^{\lin}$, instead, decreases only slowly over the whole explored range.
Correspondingly, panel~(c) shows that the optimized protocol approaches the
ideal work $w^{\id,\max}$ much faster than the linear schedule, which
yields a negative work for short strokes and remains far from the ideal
value up to $\tauun=80$. 
 
The different efficiency of the two protocols is best appreciated in the
power-law regime 
shown on a logarithmic scale in Fig.~\ref{fig:cycle_work_comparison}(b). For the linear schedule, the residual energy is well described by the Kibble--Zurek
mechanism~\cite{Zurek_PRL2005,Dziarmaga_PRL2005,Polkovnikov_PRB2005, Caneva_PRB2007}.
For small $k$, each momentum mode behaves as a two-level system that undergoes a
Landau--Zener transition near its minimum gap, which occurs close to the critical point $s^*$. Excitations are created predominantly in modes with $k$ small enough that
the passage through the minimum gap is nonadiabatic. As shown in
App.~\ref{app:kz}, summing over these modes gives
\begin{equation} \label{eq:kz_linear}
\eres^{\lin}(\tauun) \simeq \frac{1}{\pi J^*}
\sqrt{\frac{\hbar\,(g_c+J_h)}{2\,\tauun}} \;, 
\end{equation}
where $J^*=J_h s^*=g_cJ_h/(g_c+J_h)$ is the common value of the coupling and of the transverse field at the critical point, $g(s^*)=J(s^*)$, and
$(g_c+J_h)/\tauun$ is the velocity at which the linear ramp
drives the system across it. 
Eq.~\eqref{eq:kz_linear} reproduces the numerical data over the whole
explored range. 
The optimized protocol residual energy (gray dashed line), instead, follows 
\begin{equation} \label{eqn:eres_opt}
\eres^{\opt} \simeq \frac{\tau_e}{\tauun} \;,    
\end{equation}
where $\tau_e$ is a constant with dimensions of time, see Sec.~\ref{sec:power_optimization}
and App.~\ref{app:residual} for a more detailed analysis.
The data in this regime are well described by 
$\tau_e=\hbar(g_c+J_h)/(2g_cJ_h)$, which we use throughout the rest of the paper. 
 
These two power laws have a direct operational meaning. The duration
needed to reach a given residual energy $\eres$ scales as
$\tau_{\rm opt}(\eres)\propto e^{-1}_{\res}$ for the optimized protocol and as
$\tau_{\lin}(\eres)\propto e^{-2}_{\res}$ for the linear schedule, so that
\begin{equation}
    \tau_{\lin}(\eres)\propto\tau^2_{\rm opt}(\eres) \;.
\end{equation}
Hence, even before the {\em controllability threshold}, occurring around $\tau^{\crit}_N$, as discussed in Sec.~\ref{sec:power_optimization}, optimal control provides a {\em quadratic speed-up} over the linear schedule, as found in
Ref.~\cite{mbeng2019optimal} for the $J=g$ case. 
Here, with the mapping of App.~\ref{app:residual}, we extend this result
to arbitrary endpoints $(g_c,J_h)$.
In passing, we note that the suppression of excitations through optimized nonlinear passages across a quantum critical point was also investigated in Ref.~\cite{Barankov_2008}, including for the transverse-field Ising model.

\begin{figure}[t]
    \centering
    \includegraphics[width=\columnwidth]{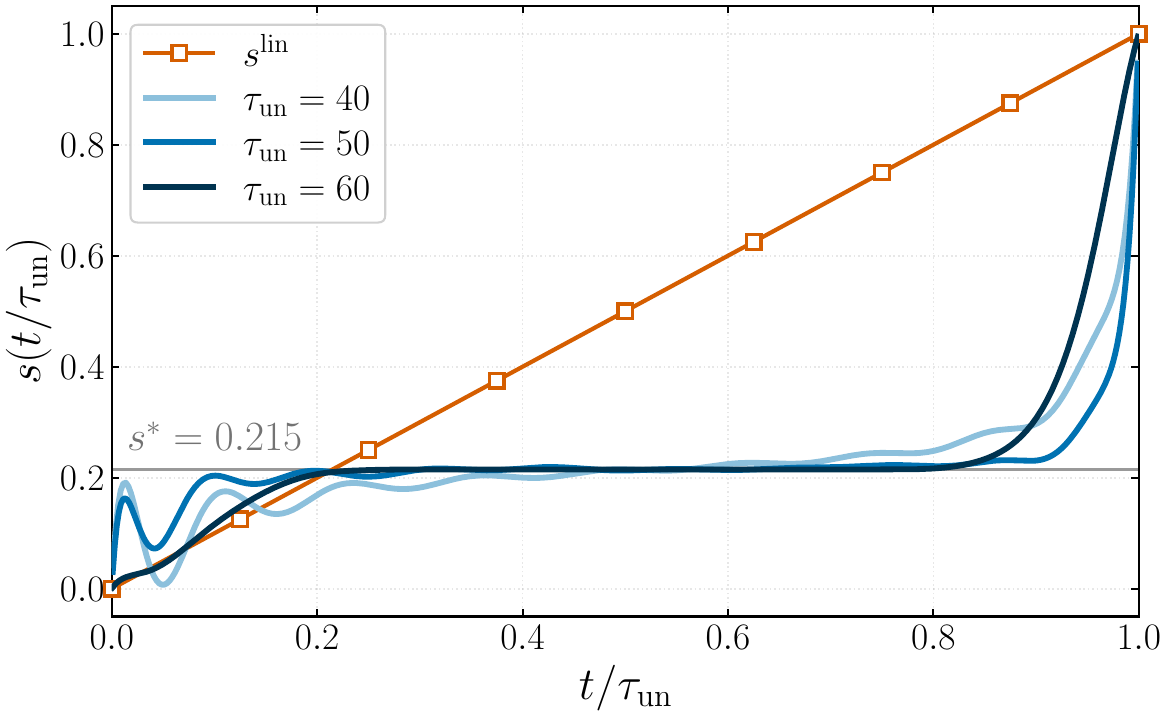}
    \caption{
    Optimized protocols for three representative values of the
    unitary-stroke duration, $\tauun=40$, $50$ and $60$, as a function of
    the normalized time $t/\tauun$, compared with the linear schedule
    $s^{\lin}(t)=t/\tauun$ (squares). 
    For each $\tauun$ we show the protocol with the lowest residual energy among all runs. The horizontal gray
    line marks the critical point $s^*\approx0.215$, around which the
    optimized protocols spend most of the evolution time. Times are in
    units of $\hbar/(k_BT_h)$.}
    \label{fig:optimized_protocols}
\end{figure}
 
In Fig.~\ref{fig:optimized_protocols} we show examples of optimized
protocols for $\tauun=40$, $50$ and $60$ $\hbar/(k_B T_h)$. Remarkably, the optimized
protocols spend most of the evolution time close to the critical point
$s^*\approx0.215$, although no information about its location is provided
to the optimizer. This behavior is reminiscent of the local adiabatic
schedule of Roland and Cerf, which slows down where the spectral gap is
minimal~\cite{Roland_PRA2002}. As discussed in Sec.~\ref{sec:power_optimization}, the
time average of the optimized schedules is close to $s^*$.
 
\subsection{$N$-dependence of the threshold time $\tau^{\thrs}_N$}
\label{sec:numerical_critical_time}
 
The results presented above show that, upon increasing the duration of the
unitary stroke, the optimized protocol progressively suppresses the final
nonadiabatic excitations and eventually approaches the ideal-work limit.
It is therefore useful to identify the characteristic control time beyond
which the optimized cycle delivers essentially the ideal work.
 
Since the relevant figure of merit for the engine is the work, rather than
the residual energy itself, we fix the accuracy threshold through the
relative work loss. From Eq.~\eqref{eq:work_logmean_range}, a residual
energy $\eres$ reduces the cycle work by a fraction
$A \eres / w^{\id,\max}$. We therefore require
\begin{equation} \label{eq:tau_star_threshold}
\frac{w^{\id,\max}-\overline{w}}{w^{\id,\max}}
= \frac{A \,\overline{e}_{\res}}{w^{\id,\max}} < \threshold \;,
\end{equation}
with $\threshold=10^{-2}$, corresponding to a typical optimized cycle delivering at least 99\% of the ideal work. 
For the parameters used here,
$A/w^{\id,\max}\approx 4.9$, and Eq.~\eqref{eq:tau_star_threshold}
corresponds to $\overline{e}_{\res} \lesssim 2\times10^{-3}$.
 
\begin{figure}[t]
    \centering
    \includegraphics[width=0.9\linewidth]{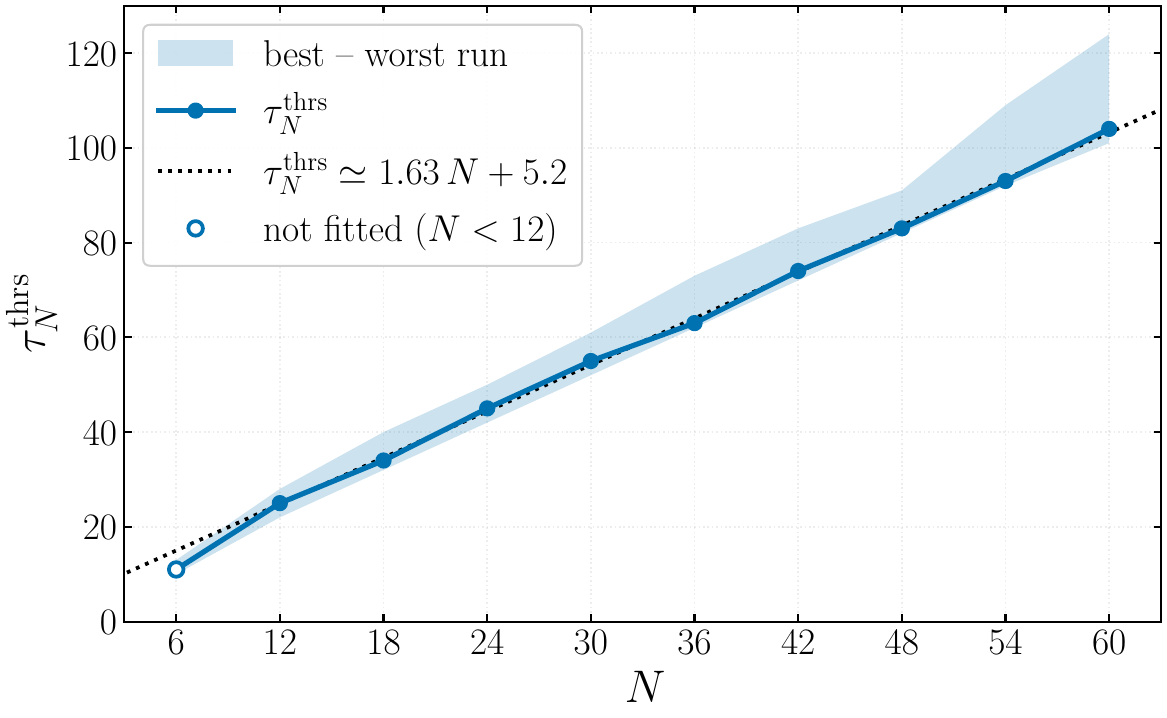}
    \caption{
    Numerical threshold time $\tau^{\thrs}_N$ as a function of the system
    size $N$. Circles mark the smallest duration from which the geometric
    mean of the residual energy over the $N_{\rm run}=10$ runs stays below
    the threshold of Eq.~\eqref{eq:tau_star_threshold}, corresponding to a
    $1\%$ loss of cycle work. The shaded band spans the values obtained by
    applying the same criterion to the best and to the worst run. The
    dotted line is the fit $\tau^{\thrs}_N\approx1.63\,N+5.2$,
    Eq.~\eqref{eq:tau_star_fit}, to the sizes $N\geq12$; the smallest size,
    $N=6$ (open circle), is excluded from the fit because of strong
    finite-size effects. Times are in units of $\hbar/(k_BT_h)$.}
    \label{fig:taucrit_numeric}
\end{figure}

For each system size $N$, we repeat the optimization procedure described in Sec.~\ref{subsec:ising_work_real}, with $N_{\rm run}$ independent runs, for increasing values of the unitary-stroke duration $\tauun$. 
We define the numerical threshold time $\tau^{\thrs}_N$ as the smallest
duration on a grid with spacing $\Delta\tauun=1\ [\hbar/(k_BT_h)]$ such that
the mean residual energy $\overline{e}_{\res}$, Eq.~\eqref{eq:eres_logmean},
satisfies Eq.~\eqref{eq:tau_star_threshold} at $\tauun=\tau^{\thrs}_N$ and at
every larger grid duration,
\begin{equation}
\tau^{\thrs}_N=\min\big\{\tauun :\ \overline{e}_{\res}(\tau')<2\times10^{-3}
\ \ \forall\,\tau'\ge\tauun\big\} \;.
\end{equation}
(Requiring the condition to hold for all subsequent
durations, rather than at a single grid point, prevents isolated favorable
optimizations from biasing the estimate.) 
Applying the same criterion to the lower and upper envelopes of the runs,
$e_{\res}^{\min}$ and $e_{\res}^{\max}$ in Eq.~\eqref{eq:eres_range},
we obtain the range of $\tauun$ within which the individual runs cross the
threshold; we use this range, shown as a shaded band in
Fig.~\ref{fig:taucrit_numeric}, as the uncertainty on $\tau^{\thrs}_N$.
 
In Fig.~\ref{fig:taucrit_numeric} we report $\tau^{\thrs}_N$ as a function of
the system size $N$. Over the investigated range, the threshold time
grows linearly with $N$. 
A fit to the sizes $N\geq12$ 
gives
\begin{equation} \label{eq:tau_star_fit}
\tau^{\thrs}_{N} \simeq \tau_1 \,N + \tau_0 \;,
\end{equation}
with $\tau_1 \approx 1.63\pm 0.02$
and $\tau_0 \approx 5.2 \pm 0.6$,
in units of $\hbar/(k_B T_h)$. 
This linear dependence of $\tau^{\thrs}_{N}$ on $N$
highlights the advantage of an optimal schedule over the linear one. 
For comparison, in a linear protocol adiabaticity requires the
evolution to be slow compared with the inverse square of the minimum
spectral gap, $\tauun\gtrsim\Delta_{\min}^{-2}$~\cite{Albash_RMP18}. 
Since at the quantum critical point $\Delta_{\min}\sim 1/N$, the corresponding control time is expected to grow as $\tauun\sim N^2$~\footnote{
Within the Kibble--Zurek picture, the linear ramp becomes adiabatic only when even the mode with the smallest gap, $k_1=\pi/N$, crosses the critical point adiabatically, i.e., when the Kibble--Zurek momentum
$k_{\rm KZ}=\big[\hbar(g_c+J_h)/(2\pi {J^*}^2\tauun)\big]^{1/2}$, see
App.~\ref{app:kz}, becomes smaller than $\pi/N$. 
This requires (see~\cite{Zurek_PRL2005,Dziarmaga_PRL2005})
\begin{equation*} \label{eq:tau_lin_adiabatic}
\tauun \gtrsim \frac{\hbar\,(g_c+J_h)}{2\pi^3 J^{*2}}\,N^2
\simeq 0.58\,N^2 \;,
\end{equation*}
in units of $\hbar/(k_BT_h)$, i.e., $\tauun\sim 520$ already for
$N=30$, well beyond the explored range. 
}. 
Optimal control, therefore, reduces the time $\tauun$ required to reach the ideal-work regime from $\mathcal{O}(N^2)$ to $\mathcal{O}(N)$.
 
\section{Optimization of the cycle power}
\label{sec:power_optimization}
So far, we have discussed several aspects of the optimization of the cycle work for an Ising Otto engine, where part of the story was to optimize the work produced at a given value of the unitary evolution time $\tauun$. Since the power of the cycle, see Eq.~\eqref{eq:power_cycle}, is the ratio between the work produced and the time it takes to execute it, it is not {\em a priori} obvious that optimizing the work is equivalent to optimizing the power: it might prove convenient, from the optimal power perspective, to execute the cycle in a shorter time, by sacrificing the work produced~\cite{Rezek2006, Insinga_2020}.
 
In this section we discuss, therefore, the optimization of the power output over a single cycle, starting from the optimization at {\em fixed endpoints} of the protocol, equal to those that maximize the work produced.
\subsection{Power optimization at fixed endpoints}
Using the expression of the work per site Eq.~\eqref{eq:A_def}, we can write the power per site in a cycle as
\begin{equation} \label{eqn:Power_vs_tauun}
\frac{\Power_{\cycle}(\tauun)}{N} = 
\frac{w^{\id,\max} -
A\, \eres(\tauun)}{2(\tauun + \tauth)} \;,
\end{equation}
where $\eres(\tauun)$ is the standard pure-state dimensionless residual energy, as obtained from Eq.~\eqref{eq:Cresen} with optimal CRAB parameters $\C$ chosen for a given value of $\tauun$, and $A$ is the thermal prefactor defined in Eq.~\eqref{eq:A_def}, which accounts for the thermal populations of the initial states of the two unitary strokes.
 
Equation~\eqref{eqn:Power_vs_tauun} makes the competition behind the
power optimization explicit: a longer unitary stroke reduces the residual
energy, and therefore increases the work in the numerator, but it also
lengthens the cycle time in the denominator. 
Where the power maximum lies depends on how fast $\eres$ decreases with $\tauun$, 
and on the thermalization time $\tauth$: 
a long $\tauth$ makes the extra duration of a
slower, more accurate unitary stroke comparatively cheap, while a short
$\tauth$ favors fast strokes even at the price of a sizable residual
energy. To make this competition quantitative we need an analytical estimate of
$\eres(\tauun)$, allowing us to optimize the cycle power in
Eq.~\eqref{eqn:Power_vs_tauun} and compare the resulting predictions with
fully numerical optimizations. 
To this end, we exploit a time-rescaling
relation, derived in App.~\ref{app:residual}, that connects the unitary
dynamics at different Hamiltonian energy scales $g_c$ and $J_h$ to a single
dimensionless reference problem with equal, unit couplings, defined
by the duality rescaling
\begin{equation}
    \tilde{H}_J=\frac{1}{J_h} H_J \;, \hspace{10mm} \tilde{H}_g=\frac{1}{g_c} H_g \;.
\end{equation}
Writing the forward Hamiltonian as
\begin{align} \label{eq:time_rescaling}
H_f(t) &= J_h s(t) \, \tilde{H}_J + g_c \big[1-s(t)\big] \tilde{H}_g  \nonumber \\
&= \Lambda(t) \, \Big( \tilde{s}(t)\,\tilde{H}_J
+ \big[1-\tilde{s}(t)\big]\,\tilde{H}_g\Big) \;,
\end{align}
with an overall energy scale $\Lambda(t)>0$ and interpolation parameter $\tilde{s}(t)$ given by
\begin{equation} \label{eq:Lambda_stilde}
   \Lambda(t) = s(t) J_h + \big[1-s(t)\big] g_c \;, \qquad  \tilde{s}(t) = \frac{J_h}{\Lambda(t)} s(t) \;,
\end{equation}
any protocol $s(t)$ between arbitrary positive endpoints $g_c$ and $J_h$ is equivalent to a protocol $\tilde{s}(\tilde{t})$ of the reference problem with $J=g=1$, associated with $\tilde{H}_J$ and $\tilde{H}_g$, evolving in the rescaled time
\begin{equation} \label{eq:t_tilde}
    d\tilde{t}= \frac{\Lambda(t)}{\hbar} \,dt
\end{equation}
with total dimensionless duration
\begin{equation} \label{eq:tau_tilde_int}
\tauuntilde=\frac{1}{\hbar} \int_0^{\tauun} \! dt\,\Lambda(t) \;.
\end{equation} 
The two protocols generate the same unitary evolution and therefore, for the corresponding pair of protocols, the same dimensionless
residual energy $\tilde{e}_{\res}(\tauuntilde) = \eres(\tauun)$.
This correspondence is exact for each individual protocol. However, the rescaled duration depends on the schedule, so protocols with the same physical duration need not have the same duration in the reference problem. Thus the mapping does not, by itself, establish an equivalence between the two fixed-duration optimization problems.

Motivated by the self-duality of the reference problem and by previous numerical studies of optimal digitized quantum annealing~\cite{mbeng_quantum_2019,mbeng2019optimal}, we adopt a balanced-schedule \textit{Ansatz} for the optimized protocols: the two normalized Hamiltonians are weighted equally on average. Equivalently, the
time average of $\tilde{s}(\tilde{t})$ equals the critical value $\tilde{s}^*=1/2$,
\begin{equation} \label{eq:mean_tildes_condition}
\frac{1}{\tauuntilde}
\int_0^{\tauuntilde}\!\! d\tilde{t}\, \tilde{s}(\tilde{t}) = \frac{1}{2} \;,
\end{equation}
Using $\tilde{s}\,d\tilde{t}=J_h\,s\,dt/\hbar$, together with the corresponding identity for $1-\tilde{s}$, this balance condition can be expressed in physical time as follows (see App.~\ref{app:residual}):
\begin{equation} \label{eq:mean_s_condition}
\frac{1}{\tauun}\int_0^{\tauun}\!\! dt\,s(t)
= \frac{g_c}{g_c+J_h} = s^* \;,
\end{equation}
where $s^*$ denotes the location of the quantum critical point along the
interpolation path. 
This assumption is supported by the optimized schedules
of Fig.~\ref{fig:optimized_protocols}, whose time averages agree with $s^*$
within a relative error of $10\%$. 
Combining the two conditions, Eqs.~\eqref{eq:mean_tildes_condition} and~\eqref{eq:mean_s_condition}, we obtain
\[
\frac{\tauuntilde}{2}
= \int_0^{\tauuntilde}d\tilde{t}\,\tilde{s}
= \frac{J_h}{\hbar}\int_0^{\tauun}dt\,s
= \frac{J_h s^*}{\hbar}\,\tauun \;,
\]
which fixes the relation between the physical duration $\tauun$ and the
dimensionless duration $\tauuntilde$ of the reference problem,
\begin{equation} \label{eq:tau_tilde}
\tauuntilde 
= \frac{2g_cJ_h}{\hbar(g_c+J_h)} \tauun\,.
\end{equation}
Under this balanced-schedule assumption, Eq.~\eqref{eq:tau_tilde} provides a common time rescaling that we use to model the optimized residual energy at different endpoints. Combining this rescaling with the approximately algebraic behavior $\tilde{e}_{\res}\simeq\tilde{\tau}_e/\tauuntilde$ observed in the reference problem below the finite-size threshold~\cite{mbeng2019optimal}, we obtain
$\eres\simeq \tilde{\tau}_e\hbar(g_c+J_h)/(2g_cJ_h\tauun)=\tau_e/\tauun$, where $\tilde{\tau}_e$ is a dimensionless constant of order one. %
 
We now approximate the optimized residual energy by a piecewise expression for two separate regimes, combining this algebraic pre-critical regime with a rapid crossover to negligible residual energy at a characteristic time
\(\tau_N^{\crit}\):
\begin{equation} \label{eqn:epsilon_res_tau}
\eres(\tauun) = 
\begin{cases} 
{\tau_e /\tauun} & \hspace{2mm} 
\mbox{for} \hspace{2mm} \tauun < \tau_N^{\crit} \,, \\
\approx 0 & \hspace{2mm} \mbox{for} \hspace{2mm} \tauun \ge \tau^{\thrs}_N \,,
\end{cases}
\end{equation}
without explicitly modeling the intermediate crossover $\tau_N^{\mathrm{crit}}\leq\tau_{\mathrm{un}}<\tau_N^{\mathrm{thrs}}$.
Here, the threshold time $\tau^{\thrs}_N=\tau_1 N + \tau_0$ is larger than $\tau^{\crit}_N$ --- see vertical lines in Fig.~\ref{fig:cycle_work_comparison} --- but the difference is small, i.e., the crossover is fast. 
The behavior of the critical time $\tau^{\crit}_N$ is similarly linear in $N$
\cite{Mbeng_PhDThesis2019New,Arezzo_inpreparation}
\begin{equation} 
\label{eq:tau_crit}
\tau^{\crit}_N=\hbar\frac{J_h+g_c}{4J_hg_c} (N-1) \;. 
\end{equation}
For $N=30$, Eq.~\eqref{eq:tau_crit} gives $\tau^{\crit}_{30}\approx 44$,
while the numerical threshold time is $\tau^{\thrs}_{30}=55$, see
Fig.~\ref{fig:taucrit_numeric}.
 
Inserting Eq.~\eqref{eqn:epsilon_res_tau} into Eq.~\eqref{eqn:Power_vs_tauun}, the power per site becomes
\begin{equation} \label{eq:piecewise_power}
\frac{\Power_{\cycle}}{N} \simeq
    \begin{cases} \displaystyle
        \frac{w^{\id,\max} - C/\tauun}{2(\tauun+\tauth)}
        & \tauun < \tau^{\crit}_N\;,
        \\[14pt]
        \displaystyle
        \frac{w^{\id,\max}}{2(\tauun + \tauth)} 
        & \tauun \geq \tau^{\thrs}_N \;,
    \end{cases}
\end{equation}
where $C=\tau_e A$, with $\tau_e$ given in Eq.~\eqref{eq:tau_e_crit_rescaling}
and $A$ in Eq.~\eqref{eq:A_def}, is a constant with dimensions of an action,
\begin{equation} \label{eq:C_def}
C=\frac{\hbar(g_c+J_h)}{2g_cJ_h}\,
\big(J_h\tanh(\beta_cg_c)+g_c\tanh(\beta_hJ_h)\big) \;.
\end{equation}
Notice that $C$ depends on the endpoints, but not on $N$.
For control times $\tauun$ beyond the finite-size threshold time, $\tauun>\tau^{\thrs}_N$, the residual energy becomes negligible and the cycle work approaches its ideal value $w^{\id,\max}$.
The expression in Eq.~\eqref{eq:piecewise_power} naturally identifies two competing operating regimes: the pre-critical regime, in which the optimum results from a trade-off between the nonadiabatic energy losses and the stroke duration $\tauun$, and the post-threshold regime, in which the ideal work is reached. 
 
Differentiating $\Power_{\cycle}(\tauun<\tau_N^{\crit})$ with respect to
$\tauun$ gives the pre-critical maximum-power optimal time
\begin{equation} \label{eq:fast_maximum_time}
\tau_{\fast} =
    \frac{C + \sqrt{C^2 + C \, w^{\id,\max} \, \tauth}}{w^{\id,\max}} \;.
\end{equation}
This result is valid provided that $\tau_{\fast}<\tau^{\crit}_N$, a condition which is eventually realized for $N\to \infty$, because $\tau^{\crit}_N$ grows linearly with $N$.
The power evaluated at this optimal point is
\begin{equation} \label{eq:fast_maximum_power}
\frac{\Power_{\fast}}{N} \equiv 
\frac{\Power_{\cycle}(\tau_{\fast})}{N} 
    =\frac{\sqrt{C^2 + C w^{\id,\max} \tauth}}{2\tau_{\fast}(\tau_{\fast}+\tauth)} 
    = \frac{C}{2\tau_{\fast}^2} \;,
\end{equation}
where the last equality, valid for any $\tauth$, follows from the
stationarity condition: at the maximum of the pre-critical branch of
Eq.~\eqref{eq:piecewise_power}, the power equals the ratio of the
derivatives of numerator and denominator, $(C/\tau_{\fast}^2)/2$. Since the constant $C$ does not depend on $N$, neither $\tau_{\fast}$ nor
$\Power_{\fast}/N$ depend on the system size: as long as
$\tau_{\fast}<\tau^{\crit}_N$, the fast-branch maximum is the same for all
sizes. 

Beyond the threshold time, $\tauun>\tau^{\thrs}_N$, 
the work is approximately the ideal one $w^{\id,\max}$, 
and the power decreases monotonically with the unitary-stroke duration
$\tauun$.
The second candidate maximum is therefore located close to the
finite-size crossover
\begin{equation} \label{eq:slow_maximum_time}
\tau_{\slow} \simeq
    \tau^{\thrs}_N = \tau_1 N + \tau_0 \;,
\end{equation}
with a corresponding power
\begin{equation} \label{eq:slow_maximum_power}
\frac{\Power_{\slow}}{N}
\simeq \frac{w^{\id,\max}}{2(\tau^{\thrs}_N+\tauth)} \;.
\end{equation}
This provides an estimate of the slow-branch maximum from the near-ideal-work regime, while its precise location within the crossover should be determined numerically.
The maximum-power operating point is determined by the competition
between these two branches. In particular, changing the thermalization
time can change which of the two local maxima provides the largest
cycle power. %
Observe that, while $\Power_{\fast}/N$ is independent of $N$, $\Power_{\slow}/N$ decreases as $N$ increases because of the increasing $\tau^{\thrs}_N$ in the denominator. 
Hence, we predict that the maximum at $\tauun=\tau_{\fast}$
will eventually dominate for $N\to\infty$. The crossover size $N_\times$
at which the global maximum moves from the slow to the fast branch follows
from imposing $\Power_{\slow}=\Power_{\fast}$, i.e.,
\begin{equation} \label{eq:N_cross}
\tau^{\thrs}_{N_\times} + \tauth = \frac{w^{\id,\max}}{2\Power_{\fast}/N}
\;.
\end{equation}
The dependence on thermalization time $\tauth$ is weak in the range of interest~\footnote{
According to
Eqs.~\eqref{eq:fast_maximum_time}, \eqref{eq:fast_maximum_power}
and~\eqref{eq:N_cross}, with the fit of Eq.~\eqref{eq:tau_star_fit},
increasing $2\tauth$ from $0$ to $20\,\hbar/(k_BT_h)$ shifts $\tau_{\fast}$
from $29.5$ to $33.8\,\hbar/(k_BT_h)$, lowers $\Power_{\fast}/N$ from
$1.54\times10^{-3}$ to $1.17\times10^{-3}\,(k_BT_h)^2/\hbar$, and moves the
crossover from $N_\times\approx33$ to $N_\times\approx38$.
For $\tauth\gg C/w^{\id,\max}$ (see Eq.~\eqref{eq:fast_maximum_time}), one finds
$\tau_{\fast}\simeq\sqrt{C\tauth/w^{\id,\max}}$ and the crossover condition
reduces to $\tau^{\thrs}_{N_\times}\simeq2\tau_{\fast}$, so that $N_\times$ grows
as $\sqrt{\tauth}$. The qualitative picture, with a fast branch that
dominates for $N>N_\times$, is therefore unchanged.
}.

\subsection{Numerical results for maximum power at fixed endpoints}
\label{subsec:power_numerics}
We now compare these predictions with the power of the optimized protocols
of Sec.~\ref{subsec:ising_work_real}, keeping the work-optimal endpoints
$g_c\approx 0.210\, k_B T_h$, $J_h \approx 0.767\,  k_B T_h$ and $T_c/T_h=0.1$.
Following Ref.~\cite{Arezzomanybodyengines}, we fix the total
thermalization time to $2\tauth=5$, in units of $\hbar/(k_BT_h)$,
independently of system size, assuming that each thermalization stroke brings the working medium to its equilibrium state. The resulting size dependence of the maximum power is therefore conditional on this assumption.
For these endpoints, 
$C\approx2.67\,\hbar$ from Eq.~\eqref{eq:C_def}, and the analytic estimates
of Eqs.~\eqref{eq:fast_maximum_time} and~\eqref{eq:fast_maximum_power}
give
\begin{equation} \label{eq:fast_numbers}
  \tau_{\fast} \approx 30.7\,\frac{\hbar}{k_BT_h}\,, \quad
    \frac{\Power_{\fast}}{N} \approx 1.42\times10^{-3}\,
\frac{(k_BT_h)^2}{\hbar}
 .
\end{equation}
For each size $N$ and duration $\tauun$, the power is computed from the
same runs used for the cycle work in Sec.~\ref{subsec:ising_work_real}:
the typical power is obtained from the geometric mean of the residual energy over the $N_{\rm run}=10$ runs, and its range from the best and
the worst run, see Eq.~\eqref{eq:work_logmean_range}.
 
\begin{figure}[t]
    \centering
    \includegraphics[width=\columnwidth]{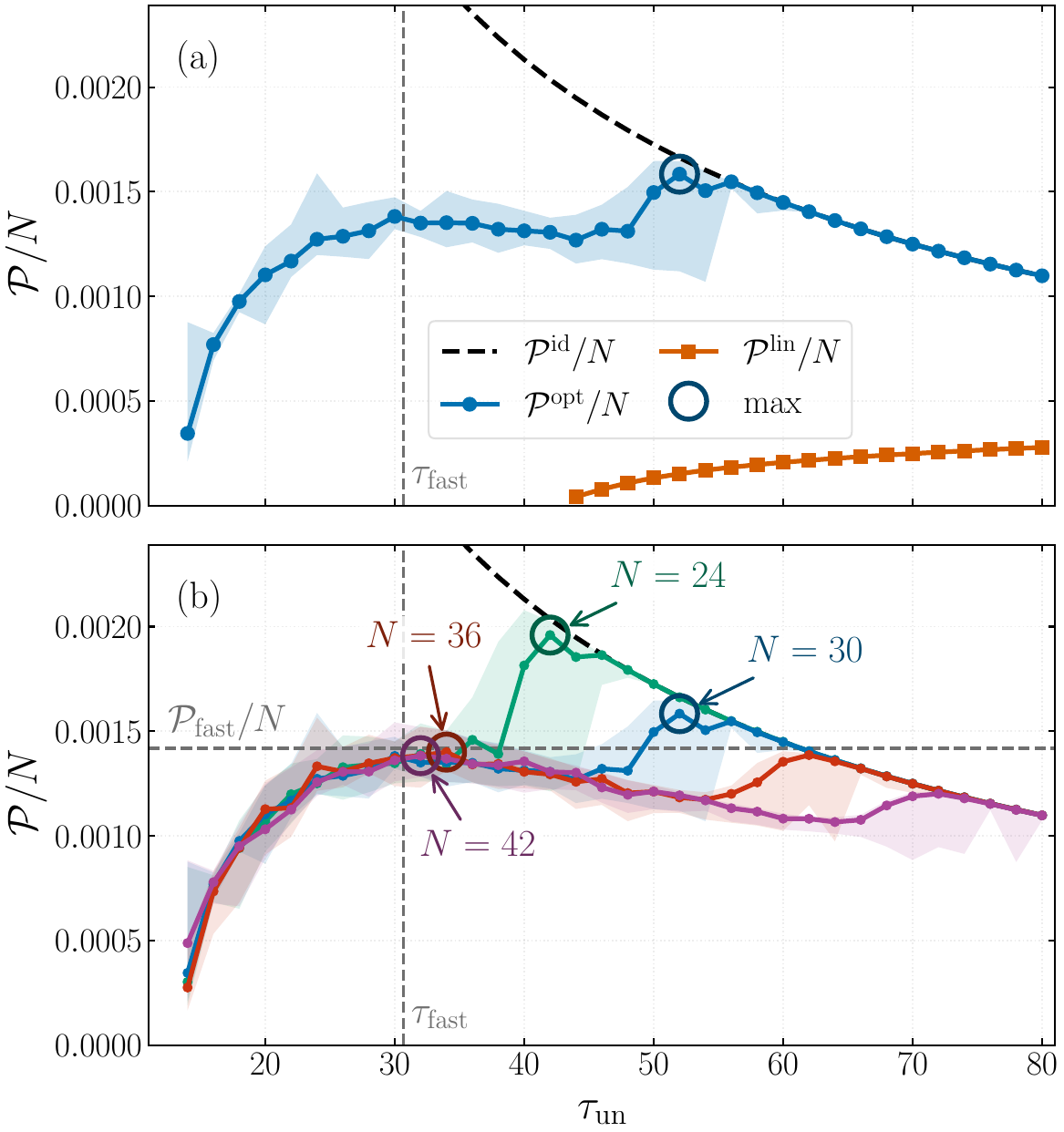}
    \caption{
    Cycle power per site as a function of the duration $\tauun$ of each
    unitary stroke, for $2\tauth=5$.
    (a) $N=30$: optimized protocol (circles), linear schedule (squares),
    and ideal power in Eq.~\eqref{eqn:Pideal} (black dashed line).
    For the optimized protocol, circles show the power obtained
    from the geometric mean of the residual energy over the $N_{\rm run}=10$
    runs, and the shaded band spans the best and the worst run; the large
    open circle marks the maximum.
    (b) Optimized power for $N=24$, $30$, $36$, and $42$, with the same
    statistics. Large open circles mark the maximum of each curve, and the
    horizontal gray dashed line is $\Power_{\fast}/N$,
    Eq.~\eqref{eq:fast_maximum_power}.
    In both panels, the vertical gray dashed line marks $\tau_{\fast}$,
    Eq.~\eqref{eq:fast_maximum_time}. For $N=24$ and $30$ the maximum lies
    on the slow branch, just below $\tau^{\thrs}_N$; for $N=36$ and $42$ it lies
    on the fast branch, close to $\tau_{\fast}$.
    Times are in units of $\hbar/(k_BT_h)$ and powers in units of
    $(k_BT_h)^2/\hbar$.
    }
    \label{fig:cycle_power_comparison}
\end{figure}

Figure~\ref{fig:cycle_power_comparison}(a) shows the cycle power per site
as a function of $\tauun$ for $N=30$. The optimized protocol yields a much
larger power than the linear schedule over the whole explored range. Its
power displays two local maxima. The first lies close to $\tau_{\fast}$,
where the work is still reduced by a finite residual energy; the second
lies just below $\tau^{\thrs}_N$, where the optimized protocol starts to deliver
the ideal work and the power essentially joins the ideal curve
\begin{equation} \label{eqn:Pideal}
    \frac{\Power^{\id}}{N} = \frac{w^{\id,\max}}{2(\tauun+\tauth)} \;,
\end{equation} 
which then decreases as $1/\tauun$. 
Figure~\ref{fig:cycle_power_comparison}(b) shows how the two
maxima evolve with $N$. The fast-branch part of the curves is nearly
independent of the size, as predicted by
Eq.~\eqref{eq:fast_maximum_power}, while the slow maximum follows
$\tau^{\thrs}_N$ to longer durations and its height decreases. As a result, the
global maximum lies on the slow branch for $N=24$ and $30$, and on the
fast branch for $N=36$ and $42$.
 
\begin{figure}[!tbp]
    \centering
    \includegraphics[width=\columnwidth]{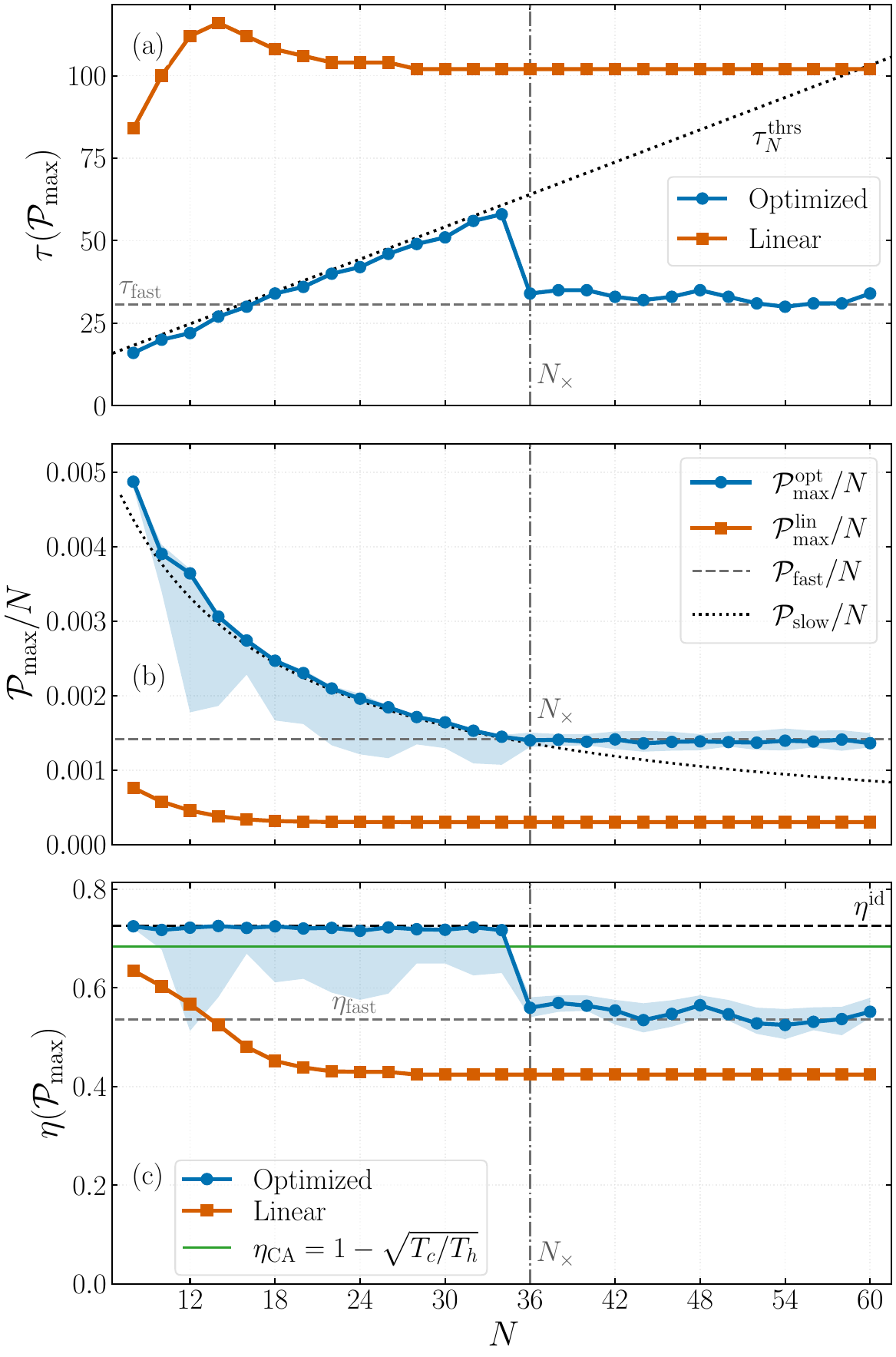}
    \caption{
    Maximum cycle power as a function of the system size $N$, for
$2\tau_{\rm th}=5$.
(a) Unitary-stroke duration $\tau(\mathcal{P}_{\max})$ at which the
power is maximal, for the optimized (circles) and linear (squares)
protocols. The black dotted line is the fit of the threshold
time, $\tau^{\thrs}_N$, Eq.~\eqref{eq:tau_star_fit}, and the
gray dashed line is $\tau_{\fast}$ of
Eq.~\eqref{eq:fast_maximum_time}.
The durations are scanned with step $\Delta\tauun=1$; since the fast-branch
maximum is flat, the values of $\tau(\mathcal{P}_{\max})$ are only indicative.
(b) Maximum power per site. For the optimized protocol, circles show
the maximum of the power obtained from the geometric mean of the
residual energy over the $N_{\rm run}=10$ runs, and the shaded band
spans the best and worst run at the same duration.
The gray dashed and black dotted
lines are the analytic fast- and slow-branch powers,
Eqs.~\eqref{eq:fast_maximum_power} and~\eqref{eq:slow_maximum_power},
the latter evaluated with the fit of $\tau^{\thrs}_N$.
(c) Efficiency $\eta=W_{\cycle}/Q_h$, Eq.~\eqref{eq:eta_eres}, at the
duration of maximum power, for the optimized (circles) and linear
(squares) protocols; the shaded band spans the best and worst run at
the same duration. The black dashed line is the ideal Otto
efficiency $\eta^{\id}=1-g_c/J_h$, reached when $\eres=0$, the gray
dashed line is the fast-branch value $\eta_{\fast}=\eta(\tau_{\fast})$,
and the green solid line is the Curzon--Ahlborn
efficiency~\cite{curzon1975efficiency} $\eta_{\rm CA}=1-\sqrt{T_c/T_h}$,
shown for reference.
In all panels, the dash-dotted line labeled $N_\times$ marks
$N_\times=36$, the smallest size whose maximum power lies on the fast
branch. This is consistent with the analytic estimate
$N_\times\approx34.4$ of Eq.~\eqref{eq:N_cross}.
Times are in units of $\hbar/(k_BT_h)$ and powers in units of
$(k_BT_h)^2/\hbar$.}
    \label{fig:power_vs_N}
\end{figure}
 
Figure~\ref{fig:power_vs_N} summarizes the maximum-power operating point
as a function of $N$. For $N\leq34$ the maximum lies on the slow branch:
its duration, panel~(a), grows linearly with $N$ just below $\tau^{\thrs}_N$,
and the maximum power, panel~(b), decreases with $N$ following
$\Power_{\slow}/N$ of Eq.~\eqref{eq:slow_maximum_power}. From $N=36$ on,
the maximum jumps to the fast branch: its duration becomes independent of
$N$ and close to $\tau_{\fast}$, and the maximum power per site saturates
to a plateau,
\begin{equation} \label{eq:power_plateau}
\left.\frac{\Power_{\max}}{N}\right|_{\rm num}\approx1.39\times10^{-3}\,
\frac{(k_BT_h)^2}{\hbar} \;,
\end{equation}
within $3\%$ of the analytic estimate of Eq.~\eqref{eq:fast_numbers}, which
contains no parameter fitted on the power.
The numerical crossover, between $N=34$ and $N=36$, agrees with the
analytic estimate $N_\times\approx34.4$ obtained from
Eq.~\eqref{eq:N_cross} with the fit of Eq.~\eqref{eq:tau_star_fit}. On the
fast branch, the engine therefore delivers a power that grows {\em extensively}
with $N$, at the price of a finite residual energy: at
$\tau_{\fast}$ the cycle work per site is about half of the ideal one (see Fig.~\ref{fig:cycle_work_comparison}, panel~(c) at $\tauun \approx 31 \hbar/(k_B T_h)$).
 
The linear schedule behaves differently. For $N\geq28$ its maximum power per site is
also independent of $N$, but it is reached at much longer durations,
$\tauun\approx10^2$, and is about $4.6$ times smaller than the fast-branch plateau of
Eq.~\eqref{eq:power_plateau}, $\Power^{\lin}_{\max}/N\approx3.0\times10^{-4}$.
These numerical values follow from the Kibble--Zurek residual energy,
Eq.~\eqref{eq:kz_linear}, which we write as
$\eres^{\lin}=B/\sqrt{\tauun}$. Inserting it into the definition of power,
Eq.~\eqref{eqn:Power_vs_tauun},
we obtain analytically that the maximum is reached for $\tauun\approx98$ and
$\Power^{\lin}_{\max}/N\approx3.05\times10^{-4}$, compatible with the numerical values.
Since the Kibble--Zurek residual energy per site is independent of $N$, the
power of the linear schedule is also extensive~\cite{revathy2020universal}; this also accounts for the behavior observed in
Ref.~\cite{Arezzomanybodyengines} for linear ramps of the transverse field
alone, where the maximum power per site occurred at a duration essentially
independent of the system size.
The gain brought by optimal control is thus a size-independent factor in
the power density, together with a change in the scaling of the time
needed to reach the ideal work, from $\mathcal{O}(N^2)$ to
$\mathcal{O}(N)$.
 
A complementary figure of merit is the efficiency at maximum power,
$\eta=W_{\cycle}/Q_h$, with $Q_h$ the heat absorbed from the hot bath,
Eq.~\eqref{eq:Qh}. For a non-ideal forward stroke, the energy of the
state reaching the hot bath exceeds the adiabatic one by
$NJ_h\tanh(\beta_cg_c)\,\eres$, see App.~\ref{app:thermoising}, so that
the heat absorbed per site is
\begin{equation} \label{eq:Qh_eres}
\frac{Q_h}{N} = q^{\id} - J_h\tanh(\beta_cg_c)\,\eres \;,
\end{equation}
where
\begin{equation}
q^{\id} = J_h\big[\tanh(\beta_cg_c)-\tanh(\beta_hJ_h)\big]
\end{equation}
is its ideal value. Together with Eq.~\eqref{eq:A_def}, this gives
\begin{equation} \label{eq:eta_eres}
\eta(\tauun) =
\frac{w^{\id,\max} - A\,\eres(\tauun)}
{q^{\id} - J_h\tanh(\beta_cg_c)\,\eres(\tauun)} \;,
\end{equation}
which reduces to the ideal Otto efficiency
$\eta^{\id}=w^{\id,\max}/q^{\id}=1-g_c/J_h\approx0.726$ for $\eres=0$, i.e., $\eta^{\id}_{\max W}$ of Sec.~\ref{sec:ideal_work}. Since
$A/w^{\id,\max}\approx4.9$ is larger than
$J_h\tanh(\beta_cg_c)/q^{\id}\approx3.0$, residual excitations
reduce the work more than the absorbed heat, and $\eta$ decreases as
$\eres$ grows. Figure~\ref{fig:power_vs_N}(c) shows $\eta$ at the
maximum-power point. On the slow branch, $N\leq34$, the engine works
essentially at the ideal efficiency, $\eta\approx0.72$--$0.73$. On the fast
branch, $\eta\approx0.54$, in agreement with
$\eta_{\fast}\equiv\eta(\tau_{\fast})\approx0.536$ obtained from
Eq.~\eqref{eq:eta_eres} with $\eres=\tau_e/\tau_{\fast}$. The transition
to the fast branch thus trades about one quarter of the efficiency for a
power that no longer decreases with $N$. The linear schedule works at
$\eta\approx0.42$ for the largest sizes, again consistent with the
Kibble--Zurek estimate, $\eta\approx0.41$.
 
For reference, Fig.~\ref{fig:power_vs_N}(c) also shows the Curzon--Ahlborn
efficiency $\eta_{\rm CA}=1-\sqrt{T_c/T_h}\approx0.684$, obtained at maximum power for the endoreversible Carnot model
with Newtonian heat transfer~\cite{curzon1975efficiency} and also recovered for
low-dissipation Carnot engines with symmetric dissipation~\cite{esposito2010efficiency}.
The Curzon--Ahlborn efficiency is not a universal bound on efficiency at maximum power. In our model, thermalization is assumed to be complete within a prescribed time $\tauth$, and nonadiabatic excitations during the unitary strokes are the only source of work loss relative to the adiabatic, fully thermalized Otto-cycle benchmark. Complete thermalization does not, however, make the isochores reversible: entropy is generally produced during heat exchange even when the unitary strokes are quantum adiabatic.
For the parameters considered, the slow-branch
efficiency exceeds $\eta_{\rm CA}$, while the fast-branch one lies below
it.
 
\subsection{Endpoint optimization for maximum power}
\label{subsec:power_endpoints}
\begin{figure}[t]
    \centering
    \includegraphics[width=\columnwidth
    ]{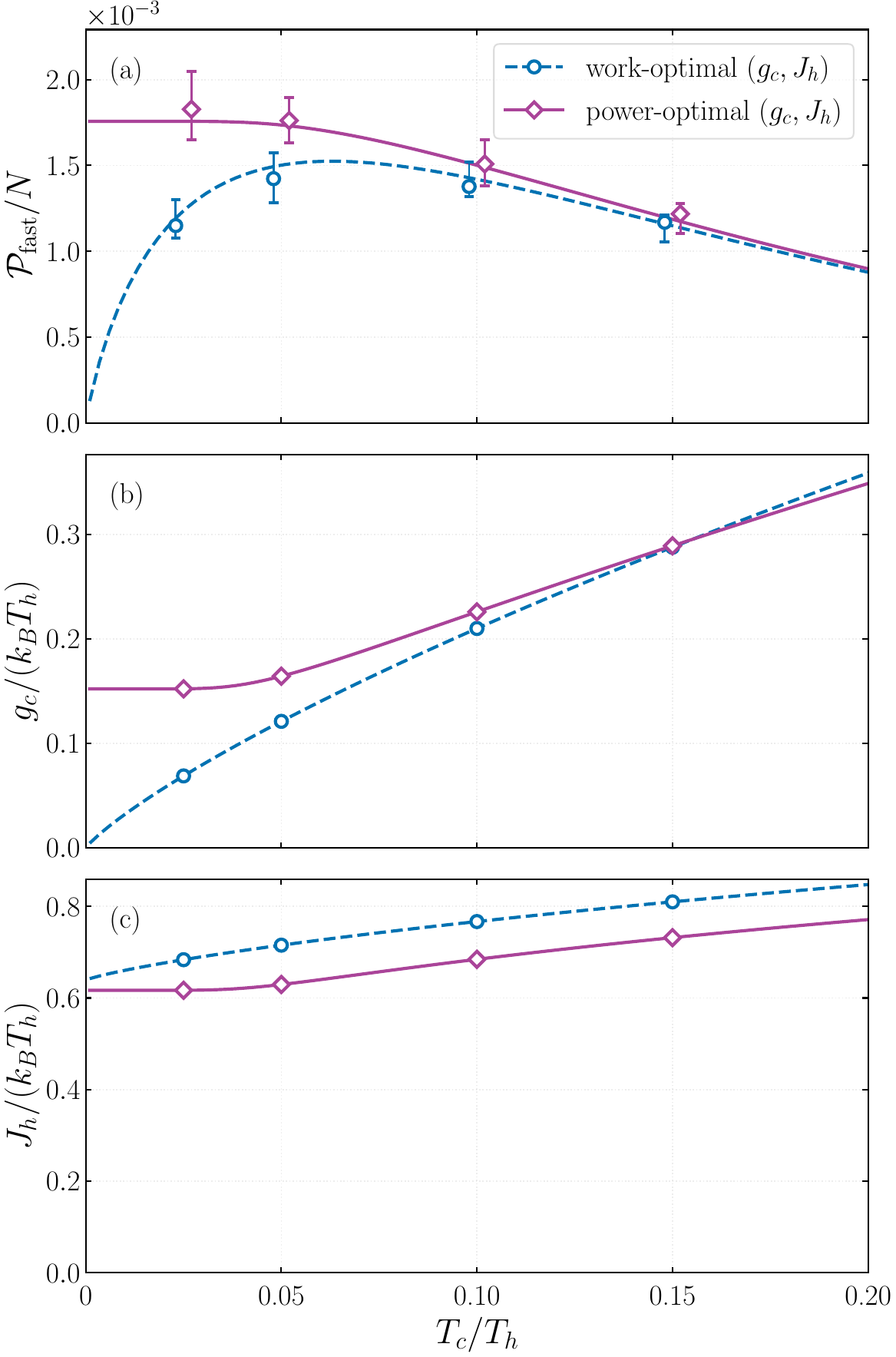}
    \caption{
        Choice of the endpoints for maximum fast-branch power, as a
        function of the temperature ratio $T_c/T_h$, for $2\tauth=5$.
        (a) Fast-branch power per site,
        Eq.~\eqref{eq:fast_power_endpoints}, for the work-optimal
        endpoints $(g_c,J_h)=(\epsilon^{\max,c},\epsilon^{\max,h})$ of Fig.~\ref{fig:ideal_work_vs_temperature_ratio} (dashed line), used in Figs.~\ref{fig:cycle_work_comparison}--\ref{fig:power_vs_N} at $T_c/T_h=0.1$, and for the endpoints that maximize
        $\Power_{\fast}/N$ (solid line). Symbols: fast-branch
        power calculated with the optimized protocol for $N=50$ at the work-optimal
        (circles) and power-optimal (diamonds) endpoints. The data are obtained from the
        geometric mean of the residual energy over the $N_{\rm run}=10$
        runs, with error bars spanning the best and the worst run at the
        same duration, as in Fig.~\ref{fig:power_vs_N}. The points of the
        two families are slightly offset horizontally for readability.
        (b), (c) Corresponding values of $g_c$ and $J_h$; symbols mark the
        endpoints used in the simulations of panel (a).
        Direct optimization of the fast-branch power yields a larger power
        at the cost of a lower ideal work per cycle.
        Energies are in units of $k_BT_h$ and powers in units of
        $(k_BT_h)^2/\hbar$.
    }
 
    \label{fig:fast_power_optimization}
\end{figure}

For $N>N_\times$ the maximum power lies on the fast branch, whose height,
Eq.~\eqref{eq:fast_maximum_power}, does not depend on $N$ and depends on
the endpoints only through the ideal work per site $w^{\id}$ and $C$. The work-optimal
endpoints of Sec.~\ref{sec:ideal_work}, used so far, are therefore not
necessarily the best choice for the power: endpoints with a slightly
smaller ideal work may give a faster crossing of the critical point, i.e.,
a smaller $C$ and a shorter $\tau_{\fast}$, and hence a larger power.
Since the fast branch is the relevant one in the large-$N$ limit, we now
optimize the endpoints directly for the fast-branch power. Inserting
Eq.~\eqref{eq:fast_maximum_time} into Eq.~\eqref{eq:fast_maximum_power},
we obtain
\begin{equation} \label{eq:fast_power_endpoints}
\frac{\Power_{\fast}}{N} =
    \frac{C \left(w^{\id}\right)^2}{2\left(C +
        \sqrt{C^2 + C w^{\id} \tauth}\right)^2} \;,
\end{equation}
with
\begin{equation} \label{eq:C_endpoints}
C = \frac{\hbar(g_c+J_h)}{2g_cJ_h} 
    \big[J_h\tanh(\beta_c g_c) + g_c\tanh(\beta_h J_h)\big] 
\end{equation}
and
\begin{equation} \label{eq:wid_endpoints}
w^{\id} = (J_h-g_c) \big[\tanh(\beta_c g_c)-\tanh(\beta_hJ_h) \big] \;,
\end{equation}
which coincides with $w^{\id,\max}$ at the work-optimal endpoints.
At fixed $T_h$ and $\tauth$, we maximize Eq.~\eqref{eq:fast_power_endpoints}
with respect to $g_c$ and $J_h$ for different values of the temperature
ratio $T_c/T_h$. 
In this optimization, the thermalization time is held fixed as the endpoint Hamiltonians are varied. The resulting endpoints maximize the model fast-branch power under this constraint, while endpoint-dependent relaxation times could shift the optimum.
In Fig.~\ref{fig:fast_power_optimization} we compare results of the optimization with the fast-branch power calculated from work-optimal endpoints. 
The power-optimal endpoints have a smaller $J_h$ and, at low temperature
ratios, a larger $g_c$. At low $T_c/T_h$ this increases the coupling at the
critical point, $J^*=g_cJ_h/(g_c+J_h)$, and therefore reduces the time
scale $\tau_e\propto 1/J^*$, %
at the cost of a lower ideal work. 
The resulting power gain is largest at low temperature
ratios, about $42\%$ for $T_c/T_h=0.025$ and $16\%$ for $T_c/T_h=0.05$,
with a reduction of the ideal work by $14\%$ and $7\%$, respectively,
and becomes small for $T_c/T_h\gtrsim0.1$ ($6\%$ at $T_c/T_h=0.1$, where
the ideal work decreases by $3\%$).
 
To test these predictions, we repeated the optimization of the unitary
strokes for $N=50$, beyond $N_\times$, at both sets of endpoints and
for $T_c/T_h=0.025$, $0.05$, $0.1$, and $0.15$, scanning $\tauun$ around
the corresponding $\tau_{\fast}$. 
The maximum fast-branch power, with the
same statistics as in Fig.~\ref{fig:power_vs_N}, is shown by the symbols in Fig.~\ref{fig:fast_power_optimization}(a). For all temperature
ratios and both choices of the endpoints, the analytic curve lies within
the range spanned by the runs, and the geometric mean differs from the
analytic estimate by at most $7\%$. The numerical simulations confirm the gain obtained by re-optimizing the endpoints. 
This is a genuine test of the model: none of the analytic curves contains
parameters fitted to these data.
At the lowest temperature ratio the numerical gain is even larger
than predicted, about $60\%$ at $T_c/T_h=0.025$, compared with the analytic
$42\%$.

\section{Conclusions}
\label{sec:conclusions}
 
In this work, we investigated the finite-time performance of a quantum Otto engine based on a transverse-field Ising chain, choosing working points connected by unitary strokes that cross the quantum critical point. The free-fermion representation allowed us to combine analytical thermodynamic results with numerical quantum optimal control, systematically investigating the dependence of work and power extraction on the driving time and system size.
 
We first established the maximum ideal-work benchmark for a free-fermion Otto engine operating between two fixed reservoir temperatures, identifying the optimal single-particle energy scales. For a collection of independent quasiparticle modes, the resulting maximum work scales extensively with the system size.
 
We then investigated finite-time work extraction, assuming complete thermalization with the reservoirs and optimizing the unitary strokes with the CRAB technique along a fixed straight path crossing the quantum critical point. For the chosen Ising endpoints, the work losses can be expressed in terms of a single residual energy, allowing us to reduce the optimization of the entire cycle to that of a single unitary stroke. We found that quantum optimal control approaches the ideal work output within 1\% on a characteristic time scale growing approximately linearly with system size, \(\tau^{\thrs}_N\simeq \tau_1 N+\tau_0\).
This contrasts with the quadratic scaling required by linear driving, consistent with the Kibble--Zurek mechanism.
 
To investigate the cycle power, we developed an analytical  model of the optimized nonadiabatic work losses. The key ingredient is a time-rescaling relation that connects protocols with different Hamiltonian energy scales to a single reference problem. Combined with an assumption on the time-averaged structure of the optimized schedules, this relation allows us to predict the residual energy as a function of the stroke duration and the Hamiltonian parameters. The resulting model identifies two competing maximum-power regimes: a slow branch approaching the ideal work output, whose optimal duration grows linearly with system size, and a fast branch that sacrifices part of the extracted work in exchange for shorter strokes. The latter exhibits a size-independent optimal duration and an extensive power output. The model further predicts a crossover between the two regimes at
$N_\times\approx 34.4$ for the work-optimal endpoints at $T_c/T_h=0.1$ and
$2\tauth=5\,\hbar/(k_BT_h)$, in quantitative agreement with the numerical
optimal-control results.
 
Beyond the crossover, optimal control yields a maximum power per site approximately $4.6$ times larger than that obtained with linear driving. This enhancement comes with a trade-off in efficiency: for the work-optimal endpoints at $T_c/T_h=0.1$, the efficiency at maximum power decreases from the ideal Otto value $\eta\approx 0.73$ on the slow branch to $\eta\approx 0.54$ on the fast branch, while remaining above the value $\eta\approx 0.42$ obtained with linear driving. For comparison, the Curzon--Ahlborn efficiency $\eta_{\mathrm{CA}}=1-\sqrt{T_c/T_h}\approx 0.684$ lies between the efficiencies of the two optimized regimes. Although it is not a bound for our engine, this comparison highlights the different efficiency--power trade-offs arising from finite-time quantum control.
 
Finally, our analytical approach allows us to optimize the Hamiltonian energy scales directly for maximum power, rather than restricting the analysis to the endpoints that maximize ideal work. We find that the power-optimal operating conditions generally differ from the work-optimal ones, reflecting a trade-off between the ideal work output and the characteristic time scale of nonadiabatic excitations. In particular, the power-optimal endpoints favor a faster critical crossing at the expense of a lower ideal work output. This leads to a substantial power enhancement at low temperature ratios, reaching approximately $42\%$ in our analytical prediction and $60\%$ in the numerical simulations at $T_c/T_h=0.025$. The agreement with independent numerical optimal-control calculations confirms the predictive capability of our approach.
 
Our analysis assumes complete thermalization with the reservoirs in a fixed time, while the optimization focuses on the nonadiabatic work losses generated during the unitary strokes. Relaxing this assumption to account for finite-time thermalization dynamics would introduce additional constraints on the engine performance and add another source of work losses. 
 
Another natural extension is to optimize the path in Hamiltonian
parameter space, rather than only its temporal parametrization, at a
fixed overall energy scale, exploring paths that differ from the straight
one.
 
Interactions could enlarge the work resource itself. For free
fermions the ideal work is at most $W_0(e^{-1})\,k_BT_h\approx0.28\,k_BT_h$
per mode, whereas for a generic many-body spectrum it is bounded by
$N k_BT_h\ln2$, a bound approached by spectra with a unique ground state
and an exponentially degenerate excited
flat band~\cite{Allahverdyan2008,Allahverdyan2013}. Such spectra, however,
require strong interactions and lie close to first-order transitions,
where relaxation times~\cite{Allahverdyan2013} and the gaps met by the
unitary strokes~\cite{Jorg2010} scale unfavorably with system size. Whether
interactions can enhance the power, and not only the ideal work, remains
an open question.
 
Finally, extending our approach to quantum Otto refrigerators could reveal how optimal control affects the trade-off between cooling power and coefficient of performance, and whether similar competing operating regimes emerge.

\begin{acknowledgments}
G.E.S. acknowledges financial support from the QuantERA II Programme STAQS project that has received funding from the European Union's H2020 research and innovation programme under Grant Agreement No.~101017733.
\end{acknowledgments}

\appendix
 
\section{Evaluation of the ideal work}\label{app:conti_max_work}
 
In this appendix we report all the calculations referring to the results in Sec.~\ref{sec:ideal_work}. 
\subsection{Free-fermion Otto engine}
Let us maximize the single element $W_{\mu}^{\id}$ in Eq.~\eqref{eq:wk} with respect to $T_c, \,\epsilon_{\mu}^c, \, \epsilon_{\mu}^h$, fixing $T_h$. 
Hereafter, for brevity, we omit the label $\mu$ and set %
$\epsilon_{\mu}^{c,h} \equiv \epsilon^{c,h}$.
By taking a derivative of the work with respect to $T_c$ we obtain
\begin{equation}
\frac{\partial W^{\id}}{\partial T_c} = 
-\frac{1}{k_B T_c^2} 4 \epsilon^c(\epsilon^h-\epsilon^c) \frac{e^{2\beta_c\epsilon^c}}{(e^{2\beta_c\epsilon^c}+1)^2} < 0 \;,
\end{equation}
where the inequality follows from $\epsilon^h > \epsilon^c$, required to produce work. This implies that the work is a monotonically decreasing function of $T_c$ and the maximum is reached at $T_c = 0$. 
Taking a derivative with respect to $\epsilon^c$ we find:
\begin{align}
\frac{\partial W^{\id}}{\partial \epsilon^c} &= 
2f_F(2\beta_c\epsilon^c)
\bigg[
1+2\beta_c(\epsilon^h-\epsilon^c) \frac{e^{2\beta_c\epsilon^c}}{e^{2\beta_c\epsilon^c}+1} \bigg] \nonumber \\
&-2 f_F(2\beta_h\epsilon^h) \;.
\end{align}
Let us now take the limit $T_c\to 0^+$ of the previous expression, obtaining:
\begin{equation}
\lim_{T_c \to 0^+} \frac{\partial W^{\id}}{\partial \epsilon^c} 
= -2 f_F(2\beta_h\epsilon^h) < 0 \;,
\end{equation}
i.e., the work is maximized for $\epsilon^c \to 0^+$. 
Since both $\epsilon^c,T_c \to 0^+$ we need to specify the behavior of the ratio $\epsilon^c/(k_BT_c)=\beta_c\epsilon^c$. 
To maximize the single-particle work, we have to take the limit by imposing $\lim_{T_c \to 0^+} \beta_c\epsilon^c = +\infty$, i.e. $k_BT_c \ll \epsilon^c$. 
 
Let us now maximize $W^{\id}$ with respect to $\epsilon^h$:
\begin{align}
\frac{\partial W^{\id}}{\partial \epsilon^h} &= 
2f_F(2\beta_h\epsilon^h)
\bigg[
1-2\beta_h(\epsilon^h-\epsilon^c) \frac{e^{2\beta_h\epsilon^h}}{e^{2\beta_h\epsilon^h}+1} \bigg] \nonumber \\
&-2f_F(2\beta_c\epsilon^c) \;. 
\end{align}
In the limit $T_c, \epsilon^c \to 0$, with $\beta_c\epsilon^c \to + \infty$, we obtain
\begin{equation}
\lim_{\substack{T_c,\,\epsilon^c \to 0 \\ \beta_c\epsilon^c \to +\infty}}  
\frac{\partial W^{\id}}{\partial \epsilon^h} = 
\frac{2}{e^{2\beta_h\epsilon^h}+1} \bigg(1- 2\beta_h\epsilon^h \frac{e^{2\beta_h\epsilon^h}}{e^{2\beta_h\epsilon^h}+1}\bigg) \;,
\end{equation}
We want this derivative to be vanishing. By defining $x \equiv 2 \beta_h \epsilon^h$ we have
\begin{equation}
x\frac{e^x}{1 + e^x} = 1 \hspace{3mm} \longleftrightarrow \hspace{3mm}
\left(x-1\right)e^{x-1} = \frac{1}{e}\;.
\end{equation}
This expression can be cast as $ze^z =1/e$ for  $z = x-1$. We can introduce the Lambert function $W_0(x)$ defined as the inverse of $z\mapsto ze^z$, i.e., $W_0(ze^z)=z$. Consequently, $z = W_0(e^{-1})$ and therefore 
\begin{equation}
x = W_0(e^{-1}) + 1
\;.
\end{equation}
By substituting again $\epsilon^h = \frac{1}{2}k_BT_h x$, we obtain the final result
\begin{equation}
\label{eq:lambertsol}
\epsilon^{\max,h} = \frac{1}{2}k_B T_h\left(1 + W_0(e^{-1})\right) \approx 0.639\, k_B T_h\;.
\end{equation}
In the limit $T_c \to 0^+,\;\beta_c\epsilon^c \to + \infty $ we have,
\begin{equation}
W^{\id,\max} = 
\frac{2\epsilon^{\max, h}}{1 + e^{2\epsilon^{\max, h}/(k_B T_h)}}\;. 
\end{equation}
By substituting Eq.~\eqref{eq:lambertsol}, we obtain
\begin{equation}
\label{eq:lammaxwork}
W^{\id,\max} =k_B T_h \frac{1 + W_0(e^{-1})}{1 + e^{\left(1 + W_0(e^{-1})\right)}}\;.
\end{equation}
By definition we have
$W_0(e^{-1})e^{W_0(e^{-1})} = e^{-1}$. Therefore, %
$e^{1+W_0(e^{-1})} = [W_0(e^{-1})]^{-1}$. By substituting this result in the denominator of Eq.~\eqref{eq:lammaxwork} we find
\begin{equation}
\begin{aligned}
W^{\id,\max} &=
k_B T_h W_0(e^{-1})
\approx0.279 \, k_B T_h\;.
\end{aligned}
\end{equation}
 
\subsection{Fixing the temperatures}\label{app:fixing_T}
Let us extend the previous results to the case in which both $T_c$ and $T_h$ are fixed. In this case, we just need to optimize 
$W_{\mu}^{\id}$ in Eq.~\eqref{eq:wk}
with respect to $\epsilon^c$ and $\epsilon^h$.
By maximizing with respect to $\epsilon^c$ and $\epsilon^h$ we obtain two conditions:
\begin{align} \label{eqn:minimumcondition}
\beta_h \frac{e^{2\beta_h\epsilon^h}}{(e^{2\beta_h\epsilon^h}+1)^2}
&= \beta_c \frac{e^{2\beta_c\epsilon^c}}{(e^{2\beta_c\epsilon^c}+1)^2}\;,
\\
f_F(2\beta_h\epsilon^h) - f_F(2\beta_c\epsilon^c) &= 
2\beta_h (\epsilon^h-\epsilon^c) \frac{e^{2\beta_h\epsilon^h}}{(e^{2\beta_h\epsilon^h}+1)^2} \;.
\end{align}
These equations can only be solved numerically. To get an approximate
solution, let us consider the low-temperature limit $T_c\ll T_h$.
From
Eq.~\eqref{eqn:zerotemperaturesolution}, in the limit
$T_c\to0^+$ the optimal hot energy satisfies
\begin{equation}
    2\beta_h\epsilon^{\max,h}
    =
    \frac{2\epsilon^{\max,h}}{k_B T_h}
    =
    1+W_0(e^{-1}).
\label{eq:optimal_hot_energy_low_Tc}
\end{equation}
Therefore, by exploiting $e^{1+W_0(e^{-1})} = [W_0(e^{-1})]^{-1}$, we find
\begin{equation}
\frac{e^{2\beta_h\epsilon^{\max,h}}}{\left(1+e^{2\beta_h\epsilon^{\max,h}} \right)^2}
= \frac{ W_0(e^{-1}) }{ \left( 1+W_0(e^{-1}) \right)^2 }
\equiv \mathcal{K}_0 \approx 0.170 \;.
\label{eq:K0_definition}
\end{equation}
 
For $T_c/T_h\ll1$, the optimal hot energy remains close to
its zero-cold-temperature value,
\begin{equation}
    \epsilon^h
    \simeq
    \epsilon^{\max,h}.
\end{equation}
Substituting this result into the stationarity condition
in Eq.~\eqref{eqn:minimumcondition}, we obtain
\begin{equation}
    \frac{T_h}{T_c}
    \frac{
        e^{2\beta_c\epsilon^c}
    }{
        \left(
            1+e^{2\beta_c\epsilon^c}
        \right)^2
    }
    \simeq
    \mathcal{K}_0.
\label{eq:low_Tc_stationarity}
\end{equation}
In the same limit, the optimal cold energy satisfies
$\beta_c\epsilon^c\gg1$. Consequently,
\begin{equation}
    \frac{
        e^{2\beta_c\epsilon^c}
    }{
        \left(
            1+e^{2\beta_c\epsilon^c}
        \right)^2
    }
    \simeq
    e^{-2\beta_c\epsilon^c},
\end{equation}
and Eq.~\eqref{eq:low_Tc_stationarity} reduces to
\begin{equation}
    e^{-2\beta_c\epsilon^{\max,c}}
    \simeq
    \mathcal{K}_0\frac{T_c}{T_h}.
\label{eq:cold_energy_exponential}
\end{equation}
The optimal cold energy is therefore
\begin{equation}
    \epsilon^{\max,c}
    \simeq
    -\frac{k_B T_c}{2}
    \ln\left(
        \mathcal{K}_0\frac{T_c}{T_h}
    \right).
\label{eq:optimal_cold_energy_low_Tc}
\end{equation}
The approximation $\beta_c\epsilon^c\gg1$ used above is self-consistent: from
Eq.~\eqref{eq:optimal_cold_energy_low_Tc}, $\beta_c\epsilon^{\max,c}\simeq-\frac{1}{2}
\ln\big(\mathcal{K}_0\,T_c/T_h\big)$, which diverges logarithmically as
$T_c/T_h\to0^+$.
 
Finally, substituting
Eqs.~\eqref{eq:optimal_hot_energy_low_Tc} and
\eqref{eq:optimal_cold_energy_low_Tc} into the ideal-work
expression in Eq.~\eqref{eq:wk}, and retaining
the leading low-$T_c$ contributions, we find
\begin{equation}
\begin{aligned}
    W^{\id,\max}_{\mu}(T_c,T_h)
    &\simeq{}
    k_B T_h W_0(e^{-1})
    \\
    &-
    k_B T_c
    \frac{
        W_0(e^{-1})
    }{
        1+W_0(e^{-1})
    }
    \left[
        1-
        \ln\left(
            \mathcal{K}_0\frac{T_c}{T_h}
        \right)
    \right] .
\end{aligned}
\label{eq:maximum_work_low_Tc}
\end{equation}
Thus, the maximum ideal work approaches its
zero-cold-temperature value,
\begin{equation}
    \lim_{T_c/T_h\to0^+}
    W^{\id,\max}_{\mu}(T_c,T_h)
    =
    k_B T_h W_0(e^{-1}),
\end{equation}
with a leading correction proportional to
$T_c\ln(T_c/T_h)$.

\section{Thermodynamics of the unitary transformations}
\label{app:thermoising}
In this Appendix we derive the two results on which the analysis of the
cycle work in Sec.~\ref{subsec:ising_work_real} relies.
 
\textit{(i) Thermal work from pure-state residual energies.}
In Sec.~\ref{app:thermal_work}, we show that the work performed along the two unitary strokes,
Eqs.~\eqref{eqn:W_f_thermal} and~\eqref{eq:w_bwd}, which in principle requires the
unitary evolution of the mixed thermal states $\rho_c$ and $\rho_h$, can be
expressed exactly in terms of the residual energies of the pure-state
evolutions starting from the ground states $\ket{\psi_g}$ and
$\ket{\psi_J}$. 
 
\textit{(ii) Time-reversed backward stroke.}
In Sec.~\ref{app:optimization_evolution} we show that if the backward schedule retraces the forward one,
$s_b(t)=1-s(\tauun-t)$,
the forward and backward normalized residual energies coincide.
The time-reversed optimal forward
protocol is therefore also optimal for the backward stroke and a single
optimization determines the whole cycle.

\subsection{Work of the unitary strokes for thermal states}
\label{app:thermal_work}
\textbf{Forward unitary stroke.} 
We first consider the forward unitary transformation,
Eq.~\eqref{eq:w_fd}, which we rewrite here for convenience
\begin{equation}\label{eq:thermal_workf}
W_f = \Tr\left(\rho_c H_g\right) -
\Tr\left(U_f\rho_c U_f^\dagger H_J\right) \;,
\end{equation}
where $U_f$ is the forward evolution, defined in its Trotterized version in Eq.~\eqref{eq:trotterized_control_protocol}, $H_g$ and $H_J$ defined in Eq.~\eqref{eq:HgHj}.
We observe that both $H_g$ and $H_J$ conserve the fermion parity $(-1)^{\hat N_f}$, so
that the Gibbs state is block-diagonal in the even and odd sectors, which
evolve independently during the unitary strokes. 
We start by considering
$\Tr\!\left(\rho_c H_g\right)
\equiv \langle H_g\rangle_{\rho_c}$.
To evaluate this average we need to consider the Gibbs state projected onto the even parity sector.
\begin{equation}
\rho_c =
\frac{P_+ e^{-\beta_c H_g}}{Z_{c,+}} \;,
\qquad
Z_{c,+} = \Tr\!\left(P_+ e^{-\beta_c H_g}\right) \;,
\end{equation}
where
\begin{equation}
P_+ = \frac{1+(-1)^{\hat N_f}}{2}\;,
\end{equation}
is the projector on the even fermion-parity subspace. 
 
Due to the diagonal structure of the Hamiltonian $H_g$, the energy of a configuration with $m$ fermions is $E_m = g_c(2m-N)$.
The partition function is evaluated by restricting the trace to the configurations with an even number of fermions, obtaining
\begin{align}
Z_{c,+}
&= e^{\beta_c g_c N}\sum_{\substack{m=0\\ m\ {\rm even}}}^{N}
\binom{N}{m} e^{-2\beta_c g_c m} \;.
\end{align}
Using the equivalence
\begin{equation}
\frac{(1+y)^N+(1-y)^N}{2} =
\sum_{\substack{m=0\\ m\ {\rm even}}} \binom{N}{m} y^m\,,
\end{equation}
with $y = e^{-2 \beta_c g_c}$, we obtain the partition function
\begin{align}
Z_{c,+}
&=
\frac{1}{2}
\left\{ \left[2\cosh(\beta_c g_c)\right]^N 
+ \left[2\sinh(\beta_c g_c)\right]^N
\right\}\,.
\end{align}
The corresponding mean energy can be obtained as
\begin{equation}\label{eq:mean_Hg}
\langle H_g\rangle_{\rho_c}
= -\frac{\partial}{\partial\beta_c}\ln Z_{c,+}= -Ng_c
f_N (\beta_c g_c) \;,
\end{equation}
being the finite-size thermal factor $f_N(x)$ defined in Eq.~\eqref{eq:sizethermalfactor}.
By combining the above result with Eq.~\eqref{eqn:hamiltoniansdefinition2} and exploiting the fact that $H_g$ is translationally invariant we obtain
\begin{equation}\label{eq:quasi_thermal_fermions}
\left\{ \begin{array}{rcl}
\braket{\opcdag{j}\opc{j'}}_{\rho_c} &=& \delta_{jj'} \displaystyle \frac{1-f_N(\beta_c g_c)}{2} \vspace{3mm} \\
\langle \opc{j} \opc{j'}\rangle_{\rho_c} &=& \langle \opcdag{j} \opcdag{j'}\rangle_{\rho_c} = 0
\end{array} 
\right. \;.
\end{equation}

Moving to the second term of Eq.~\eqref{eq:thermal_workf}, we exploit the cyclic property of the trace to write
\begin{equation}
\Tr\left(U_f\rho_c U_f^\dagger H_J\right) %
= \Tr\left(\rho_c H^{\Heis}_J \right) = \braket{H_J^{\Heis}}_{\rho_c}\;,
\end{equation}
where $H^{\Heis}_J = U_f^\dagger H_J U_f$ is the Heisenberg representation of the Hamiltonian. 
The evolution preserves the quadratic structure of $H_J$~\cite{mbeng_quantum_2019}, therefore $H_J^{\Heis}$ can be cast in the effective quadratic form:
\begin{equation} \label{eq:fermionic_hamiltonian_app}
H^{\Heis}_J
={}
\sum_{j,j'=1}^{N} \Big\{
\mathbf{A}_{j'j}
\big(
\opcdag{j'}\opc{j}
-
\opc{j} \opcdag{j'}
\big)+
\big(
\mathbf{B}_{j'j}
\opcdag{j'}\opcdag{j}
+
\text{H.c.}
\big) \Big\},
\end{equation}
where $\mathbf{A}$ is complex Hermitian and $\mathbf{B}$ complex antisymmetric.
Using Eqs.~\eqref{eq:quasi_thermal_fermions} and \eqref{eq:fermionic_hamiltonian_app} we can write 
\begin{equation} \label{eqn:HJ_rhoc}
\langle H^{\Heis}_J \rangle_{\rho_c} 
= -\Tr \big( \mathbf{A} \big) \, f_N(\beta_c g_c) \;.
\end{equation}

A similar result can be derived by assuming the evolution starting from the ground state $\ket{\psi_g}=|0\rangle$ of $H_g$, the fermionic vacuum. 
In particular, 
we find
\begin{equation} \label{eqn:HJ_psig}
\langle H^{\Heis}_J \rangle_{\psi_g}
= -\Tr \big( \mathbf{A} \big) \;,
\end{equation}
and therefore
\begin{equation} \label{eqn:HJ_rhoc_psig}
\langle H^{\Heis}_J \rangle_{\rho_c}
= \langle H^{\Heis}_J \rangle_{\psi_g} \, f_N(\beta_c g_c) \;.
\end{equation}

In the case of a perfect adiabatic evolution, $\ket{\psi_g}$ is mapped, up to a phase, onto $\ket{\psi_J}$, ground state of $H_J$ at energy $E_0^J=-NJ_h$, giving the thermal expectation value
\begin{equation}
     \langle H^{\Heis}_J\rangle_{\rho_c}^{\id} =-NJ_hf_N(\beta_cg_c).
\end{equation}

To evaluate the expectation value of the Hamiltonian for non-perfect adiabatic evolution let us introduce the dimensionless residual forward energy
\begin{equation} \label{eq:eres_f_app}
    \eres^{f} = \frac{1}{J_hN}\left(\langle H^{\Heis}_J\rangle_{\psi_g} - E^J_0\right)\;.
\end{equation}
By adding and subtracting $\langle H^{\Heis}_J\rangle^{\id}_{\rho_c}$ in Eq.~\eqref{eq:thermal_workf} we find
\begin{align}
W_f &=  \langle H_g \rangle_{\rho_c} - \langle H^{\Heis}_J\rangle^{\id}_{\rho_c}-\left(\langle H^{\Heis}_J\rangle_{\rho_c} -\langle H^{\Heis}_J\rangle^{\id}_{\rho_c}\right)\; \nonumber \\
&=W^{\id}_f - J_hN f_N\left(\beta_c g_c\right)\eres^{f} \;,
\end{align}
where $W_f^{\id} = \langle H_g \rangle_{\rho_c} - \langle H^{\Heis}_J\rangle^{\id}_{\rho_c} = 
N(J_h - g_c) f_N(\beta_c g_c)$. 

\textbf{Backward unitary stroke.}
We now consider the backward work
\begin{equation}
W_b = \Tr\!\left(\rho_h H_J\right) -
\Tr\! 
\left( U_b\rho_h U_b^\dagger H_g \right) \;,
\end{equation}
where $U_b$ is the 
backward evolution operator. 
We can extend the previously derived results by exploiting a duality symmetry between $H_g$ and $H_J$. This is conveniently done in momentum space. 
By restricting the analysis to the even fermion-parity sector, we can use the Fourier transformation of
Ref.~\cite[Eq.~(C1)]{Arezzodigital2025} and the wave vectors
$k=\pm\frac{\pi}{N},\pm\frac{3\pi}{N},\ldots,\pm\frac{(N-1)\pi}{N}$
of Ref.~\cite[Eq.~(C2)]{Arezzodigital2025}, to write the two endpoint Hamiltonians
in the form of Ref.~\cite[Eqs.~(C4) and~(C5)]{Arezzodigital2025}, with
$h\to g_c$ and $J\to J_h$.
 
It is useful to group the modes with $k$ and $-k$. 
For each positive momentum $k>0$, we introduce the pseudospin operators
\begin{align}
\tau_k^z &= \opcdag{k} \opc{k} + \opcdag{-k} \opc{-k} -1 \;, \\
\tau_k^x &= \opcdag{k} \opcdag{-k} + \opc{-k}\opc{k} \;.
\end{align}
Within the even local-parity subspace $\mathrm{span}
\left\{ \ket{0}, \opcdag{k} \opcdag{-k}\ket{0} \right\}$,
these operators satisfy the Pauli algebra.
The singly occupied states of each $(k,-k)$ pair have zero energy for
both endpoint Hamiltonians and will be left invariant by the
transformation introduced below.
 
In terms of the pseudospin operators, the endpoint Hamiltonians take the form
\begin{subequations}
\begin{align}
H_g &= 2g_c \sum_{k>0} \tau_k^z \;, \label{eq:Hg_pseudospin} \\
H_J &= 2J_h \sum_{k>0} \left(\sin k\,\tau_k^x - \cos k\,\tau_k^z \right) \;. \label{eq:HJ_pseudospin}
\end{align}
\end{subequations}
This representation makes the unitary equivalence between $H_g/g_c$ and
$H_J/J_h$
easy to derive. Indeed, in each $(k,-k)$ block we can define
\begin{equation}
V_k = 
\cos\left(\frac{k}{2}\right) \tau_k^x +
\sin\left(\frac{k}{2}\right)\tau_k^z \;,
\end{equation}
on the even local-parity subspace, and extend $V_k$ as the
identity on the singly occupied subspace. 
Since $\tau_k^x$ and $\tau_k^z$ are Hermitian and anticommute, we immediately find
\begin{equation}
V_k^\dagger = V_k \;, \qquad
V_k^2 = \mathbb I \;,
\end{equation}
where $\mathbb I$ is the identity on the $(k,-k)$ Fock space. Moreover
\begin{equation}
V_k \tau_k^z V_k =
\sin k\,\tau_k^x - \cos k\,\tau_k^z \;.
\end{equation}
Defining the global unitary transformation
\begin{equation}
V = \prod_{k>0}V_k \;,
\end{equation}
we therefore have
\begin{equation}
V^\dagger = V \;, \qquad
V^2 = \mathbb I \;.
\end{equation}
Furthermore, $V$ preserves the total fermion parity and
exchanges the two normalized endpoint Hamiltonians:
\begin{equation} \label{eq:V_exchange}
V \frac{H_g}{g_c} V = \frac{H_J}{J_h} \;, \qquad
V \frac{H_J}{J_h} V = \frac{H_g}{g_c} \;.
\end{equation}
 
We can use this transformation to evaluate the backward work. The thermal state at the beginning of the
backward stroke is the Gibbs state restricted to the even fermion-parity
sector
\begin{equation}
\rho_h = \frac{P_+e^{-\beta_h H_J}}{Z_{h,+}} \;, \qquad
Z_{h,+} = \Tr\!\left( P_+e^{-\beta_h H_J} \right) \;.
\end{equation}
Defining
\begin{equation}
\rho_h' = V\rho_hV \;,
\end{equation}
and using Eq.~\eqref{eq:V_exchange}, we obtain
\begin{equation}
\rho_h' = \frac{P_+ e^{-\beta_h(J_h/g_c)H_g}}{Z_{h,+}} \;,
\end{equation}
having exploited the fact that $V$ preserves the fermion parity and commutes with $P_+$.
Therefore, $\rho_h'$ has exactly the same form as the thermal state
considered for the forward transformation, with the replacement
\begin{equation}
\beta_c g_c \longrightarrow \beta_h J_h \;.
\end{equation}
 
Using the invariance of the trace under unitary transformations, the
first contribution to the backward work becomes
\begin{align}
\Tr\! \left( \rho_h H_J \right) =
\Tr\! \left( \rho_h' V H_JV \right)
= \frac{J_h}{g_c} \Tr\! \left( \rho_h'H_g \right) \;.
\end{align}
Using Eq.~\eqref{eq:mean_Hg} with $\beta_c g_c\rightarrow\beta_hJ_h$, we have
\begin{equation}
\Tr\! \left( \rho_h'H_g \right)
= -Ng_c \, f_N(\beta_hJ_h) \;,
\end{equation}
and hence
\begin{equation}
\Tr\! \left( \rho_h H_J \right) = 
-NJ_h \, f_N(\beta_hJ_h) \;.
\end{equation}
 
We finally consider the second contribution
\begin{equation} 
\Tr\! \left( U_b \rho_h U_b^\dagger H_g
    \right) \equiv \langle H_g \rangle_{\tilde{\rho}_h} \;,
\end{equation}
with $\tilde{\rho}_h = U_b \rho_h U^{\dagger}_b$. 
Using the cyclic property of the trace, we obtain
\begin{equation}
\langle H_g \rangle_{\tilde{\rho}_h} =
\Tr\! \left( \rho_h U_b^\dagger H_g U_b \right) \;.
\end{equation}
Conjugating the full expression by $V$ gives
\begin{align}
\langle H_g \rangle_{\tilde{\rho}_h} &=
\Tr\!\Big[ \rho_h' 
\left( V U_b^\dagger V \right)
\left( V H_g V \right)
\left( V U_b V \right) \Big] \;.
\end{align}
Introducing $
\overline{U}_b = V \, U_b \, V \;$
and using
$
V H_g V = \frac{g_c}{J_h} H_J \;,
$
we obtain
\begin{equation}
\langle H_g \rangle_{\tilde{\rho}_h} =
\frac{g_c}{J_h} \Tr\! 
\left( \rho_h' 
\overline{U}_b^\dagger H_J
\overline{U}_b \right) \;.
\end{equation}
Since $\rho_h'$ is an even-parity thermal state of a Hamiltonian
proportional to $H_g$, and $\overline{U}_b$ is generated by
$VH_gV\propto H_J$ and $VH_JV\propto H_g$, and therefore preserves the
quadratic structure, the result derived for the forward transformation
can be directly applied, yielding
\begin{align}
\langle H_g \rangle_{\tilde{\rho}_h} &= \frac{g_c}{J_h} f_N(\beta_h J_h)
\bra{\psi_g} \overline{U}_b^\dagger H_J \overline{U}_b \ket{\psi_g} \;.
\end{align}
The ground states of the two endpoint Hamiltonians are related by
$V$. Indeed, Eq.~\eqref{eq:V_exchange} implies
\begin{equation}
V\ket{\psi_g} = e^{i\phi}\ket{\psi_J} \;,
\end{equation}
where the phase $\phi$ is irrelevant for expectation values. 
Using Eq.~\eqref{eq:V_exchange},
we finally obtain
\begin{equation}
\langle H_g \rangle_{\tilde{\rho}_h} =
f_N(\beta_hJ_h)
\bra{\psi_J} U_b^\dagger H_g U_b \ket{\psi_J} \;.
\end{equation}
 
For a perfectly adiabatic backward evolution 
$U_b^{\id} \ket{\psi_J} = e^{i \varphi}\ket{\psi_g}$,
and therefore
\begin{equation}
\langle H_g \rangle^{\rm id}_{\tilde{\rho}_h} =
-Ng_c\, f_N(\beta_hJ_h) \;.
\end{equation}
Consequently, the ideal backward work is
\begin{equation}
W_b^{\id} = -N(J_h-g_c)\, f_N(\beta_hJ_h) \;.
\end{equation}
Also for the backward stroke, we can define the residual energy
\begin{equation} \label{eq:eres_b_app}
\eres^{b} = 
\frac{1}{g_cN} \left(\langle H^{\Heis}_g\rangle_{\psi_J} - E^g_0 \right)\;,
\end{equation}
where $H^{\Heis}_g = U_b^\dagger H_g U_b$ and $E^g_0 = -g_c N$, and similarly write
\begin{equation}
W_b = W^{\id}_b -g_cN f_N(\beta_hJ_h) 
\eres^{b} \;. 
\end{equation}
If we consider the total work per site $W/N \equiv w$ we obtain
\begin{equation}
w = w^{\id}_N 
- J_h f_N(\beta_c g_c) \eres^{f} 
- g_c f_N(\beta_h J_h) \eres^{b}\;,
\end{equation}
with 
\begin{equation}
w^{\id}_N= \left(J_h-g_c\right)
\left( f_N(\beta_c g_c) - f_N(\beta_h J_h) \right) \;,
\end{equation}
which reduces to $w^{\id}$ of Eq.~\eqref{eq:wid_endpoints} for $f_N(x)\xrightarrow{N\to\infty}\tanh x$.

\subsection{Optimization of the evolution}
\label{app:optimization_evolution}
In this subsection we prove that the time-reversed optimal protocol is also optimal for the backward stroke. Moreover, we will prove that 
\begin{equation}
\eres^{b}=\eres^{f}= \frac{4}{N}\sum_{k>0}\big(1-F_k\big)\;,
\end{equation}
where $F_k$ is the ground-state fidelity of an independent two-level
problem in the momentum sector $k$. 
 
To prove this, we evaluate the forward residual energy $\eres^f$ using
the pseudospin representation of
Eqs.~\eqref{eq:Hg_pseudospin} and~\eqref{eq:HJ_pseudospin} and we consider the digitized evolution~\eqref{eq:trotterized_control_protocol}. 
In momentum space the evolution operator can be factorized over the local-parity subspace. Therefore the 
digitized evolution,
Eq.~\eqref{eq:trotterized_control_protocol}, can be written as 
\begin{equation}
U_f(\btheta) = \prod_{j=1}^{n} \text{U}^{k_j}(\btheta) \;,
\end{equation}
where $n=N/2$ and $k_j=(2j-1)\pi/N$, $j=1,\ldots,n$, are the positive
wave vectors of the even-parity sector, and $\text{U}^{k_j}(\btheta)$ acts
only on the subspace $\text{span}\{\ket{0}, \opcdag{k_j}\opcdag{-k_j}\ket{0}\}$.
The initial state, the fermionic vacuum
$\ket{\psi_g}=\bigotimes_{j}\ket{0_{k_j}}$, lies in the even local-parity
subspace of each pair, which is left invariant by $\tau^x_k$ and
$\tau^z_k$. 
In the basis
$\big\{\opcdag{k}\opcdag{-k}\ket{0},\ket{0}\big\}$ of this subspace,
$\tau^z_k\to\sigma^z$, $\tau^x_k\to\sigma^x$.  
Therefore we have that 
\begin{equation}
H_g \mapsto 2g_c\sigma^z \;, \qquad H_J \mapsto 2\,\text{A}^{k}_J \;,
\end{equation}
with
\begin{equation}
\text{A}^{k}_J = J_h
\begin{pmatrix}
-\cos k & \sin k \\
\sin k &  \cos k
\end{pmatrix} \;.
\end{equation}
Moreover, the vacuum of each pair is mapped onto the ground state of
$\sigma^z$,
\begin{equation}
\ket{0_k} \mapsto \ket{\phi_0} \;, \qquad \sigma^z \ket{\phi_0} = -\ket{\phi_0} \;.
\end{equation}
We can substitute these results into the digitized evolution operator finding
\begin{equation} \label{eq:Ukp_def}
\text{U}^{k}_p(\theta^g_p,\theta^J_p) =
e^{-2i \theta^g_p g_c \sigma^z/\hbar}
e^{-2i \theta^J_p \text{A}^{k}_J/\hbar} \;,
\end{equation}
and finally obtain
\begin{equation} \label{eq:nambutraslinvenergy}
     \langle H^{\Heis}_J \rangle_{\psi_g} =\sum_{j=1}^{n = N/2}{2\langle \phi_0 |\left(\text{U}^{k_j}(\btheta)\right)^{\dagger} \text{A}_J^{k_j}\text{U}^{k_j} (\btheta) |\phi_0\rangle } \;.
\end{equation}
Therefore the QAOA problem of a transverse-field Ising Hamiltonian reduces to the sum of $N/2$ problems on the Bloch sphere, 
each starting from the ground state of $\sigma^z$ and targeting the ground state of $\text{A}^{k_j}_J$ for all $k_j$.
The same result can be obtained with the Nambu formalism of
Ref.~\cite[App.~C]{Arezzodigital2025}.

The eigenvalues of the matrix $\text{A}^{k}_J$ are $E_\text{A}^\pm=\pm J_h$; we denote by $\ket{\phi_J^k}$ its ground state. Therefore we can decompose 
$\text{A}^{k}_J = J_h\big(\mathbb{1}-2\ket{\phi_J^k}\bra{\phi_J^k}\big)$. 
By introducing the ground-state
fidelity in the momentum sector $k$ 
\begin{equation}
F_{k}^f =
\left| \bra{\phi_J^{k}} \mathrm{U}^{k}(\btheta) \ket{\phi_0} \right|^2 \;,
\end{equation}
Eq.~\eqref{eq:nambutraslinvenergy} reduces to
$\langle H^{\Heis}_J \rangle_{\psi_g} = -NJ_h +
4J_h\sum_{j=1}^{n}\big(1-F_{k_j}^f\big)$. Substituting in the expression for the residual energy we find 
\begin{equation} \label{eq:forward_residual_fidelities}
\eres^{f}(\btheta) =
\frac{\langle H^{\Heis}_J\rangle_{\psi_g} + NJ_h}{J_hN}
= \frac{4}{N} \sum_{j=1}^{N/2} \left(1-F_{k_j}^f\right) \;.
\end{equation}

We now want to evaluate the backward residual energy
corresponding to the optimized forward protocol. 
Following the same steps that led to Eq.~\eqref{eq:trotterized_control_protocol},
with $H_\text{mix}\equiv H_J$ and $H_\text{targ}\equiv H_g$, see
Eq.~\eqref{eqn:hamiltoniansdefinition_back}, we write
\begin{equation} \label{eq:trotterized_control_protocol_back}
U_b(\balpha) = \prod_{p=1}^{P}
    e^{-\frac{i}{\hbar} H_J\,\alpha^J_p}
    e^{-\frac{i}{\hbar} H_g\, \alpha^g_p} \;,
\end{equation}
with 
$\balpha=(\alpha^g_1,\ldots,\alpha^g_P,\alpha^J_1,\ldots,\alpha^J_P)$.

In the digitized representation of
Eq.~\eqref{eq:trotterized_control_protocol_back}, the time-reversed
schedule corresponds to the QAOA angles
\begin{equation} \label{eq:backward_angles}
\alpha_p^J = \theta_{P+1-p}^J \;, \qquad
\alpha_p^g = \theta_{P+1-p}^g \;, 
\end{equation}
with $p=1,\ldots,P$.
 
We notice that, since $\sigma^z$ and $\text{A}^k_J$ are real symmetric
matrices, for any $k$ in the positive-momentum sector we have
\begin{equation}
\Big[\text{U}^{k}_p(\theta^g_p,\theta^J_p)\Big]^\transpose
= e^{-2i \theta^J_p \text{A}^{k}_J/\hbar} e^{-2i \theta^g_p g_c \sigma^z/\hbar},
\end{equation}
i.e., the transpose reverses the order of the two operators.
Consequently we can write
\begin{align} \label{eq:backward_transpose}
\mathrm{U}_b^k(\balpha)
&=
\left(\mathrm{U}^k(\btheta)\right)^{\transpose} \;,
\end{align}
and it still takes a factorized form, as for the forward protocol.
We observe that also $\ket{\psi_J}=e^{-i\phi}V\ket{\psi_g}$ lies in the even
local-parity subspace of each pair, where it reduces to
$\bigotimes_j\ket{\phi_J^{k_j}}$.
Using $g_c\sigma^z = g_c\big(\mathbb{1}-2\ket{\phi_0}\bra{\phi_0}\big)$, the
same steps as for the forward stroke give
\begin{equation} \label{eq:backward_residual_fidelity}
\eres^{b}(\balpha)
=\frac{1}{g_cN}\left(
\langle H^{\Heis}_g\rangle_{\psi_J}-E_{0}^g \right)
=
\frac{4}{N}\sum_{j=1}^n \left(1-F_{k_j}^b\right) \;,
\end{equation}
with $F_k^b =
\left| \bra{\phi_0} \mathrm{U}_b^k(\balpha) \ket{\phi_J^k} \right|^2$.

Finally, we observe that both Hamiltonians are real, hence the ground states $\ket{\phi_0}$ and $\ket{\phi_J^k}$ can
be chosen real. Moreover, by exploiting  Eq.~\eqref{eq:backward_transpose} we find
\begin{align} \label{eq:equal_mode_fidelities}
F_k^b
&=
\left| \bra{\phi_0} \left(\mathrm{U}^k(\btheta)\right)^{\transpose}
\ket{\phi_J^k} \right|^2
=
\left| \bra{\phi_J^k} \mathrm{U}^k(\btheta) \ket{\phi_0} \right|^2
= F_k^f \;,
\end{align}
i.e. the two residual energies coincide for any $\btheta$,
\begin{equation} \label{eq:eres_b_equals_f}
\eres^{b}(\balpha) = \eres^{f}(\btheta) \;.
\end{equation}
Since the map $\btheta\mapsto\balpha$ of Eq.~\eqref{eq:backward_angles} is
one-to-one, the time-reversed version of the optimal forward protocol is
also optimal for the backward stroke.
\section{Time-rescaling relation and residual energy}
\label{app:residual}
We derive here the time-rescaling relation of Sec.~\ref{sec:power_optimization},
Eqs.~\eqref{eq:time_rescaling}--\eqref{eq:tau_tilde}, and the expression
of Eq.~\eqref{eqn:epsilon_res_tau} for the residual energy.%
 
\subsection{Time rescaling relation}
By Eqs.~\eqref{eq:time_rescaling} and~\eqref{eq:Lambda_stilde}, the
forward Hamiltonian is $H_f(t)=\Lambda(t)\,\tilde{H}\big[\tilde{s}(t)\big]$,
where $\tilde{H}(\tilde{s})=\tilde{s}\,\tilde{H}_J+(1-\tilde{s})\,\tilde{H}_g$
is the annealing Hamiltonian of the reference problem with $J=g=1$. Since
$\Lambda(t)\geq\min(g_c,J_h)>0$, $\tilde{s}$ is a monotonic function of $s$
with $\tilde{s}=0$ for $s=0$ and $\tilde{s}=1$ for $s=1$, so that it
satisfies the same boundary conditions as $s$, and the dimensionless time
$\tilde{t}(t)=\hbar^{-1}\int_0^t dt'\,\Lambda(t')$ of
Eq.~\eqref{eq:t_tilde} is a monotonic function of $t$. In terms of
$\tilde t$, the Schr\"odinger equation
$i\hbar\,\partial_t\ket{\psi}=H_f(t)\ket{\psi}$ becomes
\begin{equation}
i\,\frac{d}{d\tilde{t}}\ket{\psi} =
\tilde{H}\big[\tilde{s}(\tilde{t})\big]\,\ket{\psi} \;.
\end{equation}
The physical protocol $s(t)$ of duration $\tauun$ and the reference
protocol $\tilde{s}(\tilde{t})$ of duration $\tauuntilde$,
Eq.~\eqref{eq:tau_tilde_int}, therefore generate the same unitary
evolution. They also start from the same state, since $\ket{\psi_g}$ is
the ground state of $\tilde{H}_g$, and they are measured on the same
target, since
$\eres=\big(\bra{\psi(\tauun)}\tilde{H}_J\ket{\psi(\tauun)}+N\big)/N$
involves only $\tilde{H}_J$. Hence, exactly,
\begin{equation} \label{eq:eres_equal}
\eres(\tauun) = \tilde{e}_{\res}(\tauuntilde) \;.
\end{equation}
 
The relation between $\tauun$ and $\tauuntilde$ depends on the protocol.
Since $\tilde{s}\,\Lambda=s\,J_h$ and $(1-\tilde{s})\,\Lambda=(1-s)\,g_c$,
the times spent by the reference protocol on the two terms of
$\tilde{H}(\tilde{s})$ are
\begin{equation} \label{eq:time_split}
\begin{aligned}
\int_0^{\tauuntilde}\! d\tilde{t}\;\tilde{s}
&= \frac{J_h}{\hbar}\int_0^{\tauun}\! dt\; s \;, \\
\int_0^{\tauuntilde}\! d\tilde{t}\;(1-\tilde{s})
&= \frac{g_c}{\hbar}\int_0^{\tauun}\! dt\; (1-s) \;.
\end{aligned}
\end{equation}
Summing the two lines, and denoting by
$\langle s\rangle=\tauun^{-1}\int_0^{\tauun}dt\,s(t)$ the time average of
the physical schedule, we obtain
\begin{equation} \label{eq:tau_tilde_general}
\tauuntilde = \frac{\tauun}{\hbar}
\Big[J_h\langle s\rangle + g_c\big(1-\langle s\rangle\big)\Big] \;,
\end{equation}
which depends on the protocol only through $\langle s\rangle$.
 
The two dimensionless Hamiltonians are dual: the unitary $V$ of
App.~\ref{app:thermoising} exchanges them, $V\tilde{H}_gV=\tilde{H}_J$,
Eq.~\eqref{eq:V_exchange}. The {\em Ansatz} of
Eq.~\eqref{eq:mean_tildes_condition}, $\langle\tilde s\rangle=1/2$, states
that the optimized reference protocol spends on average the same time on
the two terms, so that the left-hand sides of the two lines of
Eq.~\eqref{eq:time_split} are equal. Equating the right-hand sides,
$J_h\langle s\rangle=g_c\big(1-\langle s\rangle\big)$, gives
Eq.~\eqref{eq:mean_s_condition},
\begin{equation} \label{eq:mean_s_app}
\langle s\rangle = \frac{g_c}{g_c+J_h} = s^* \;.
\end{equation}
Under the balanced-schedule assumption, inserting it into Eq.~\eqref{eq:tau_tilde_general} gives
Eq.~\eqref{eq:tau_tilde},
\begin{equation} \label{eq:bare_physical_time_relation}
\tauuntilde = \frac{2J_hs^*}{\hbar}\,\tauun
= \frac{2J^*}{\hbar}\,\tauun \;,
\end{equation}
where $J^*=J_hs^*=g_cJ_h/(g_c+J_h)$ is the common value of the coupling
and of the transverse field at the critical point. The physical duration is therefore measured in units of $\hbar/(2J^*)$, set by the coupling at the critical point.

\subsection{Residual energy}
\label{app:residual_energy}
Below the finite-size critical time, the optimized residual energy of the
reference problem displays the approximate algebraic
behavior~\cite{mbeng2019optimal}
\begin{equation} \label{eq:bare_residual_scaling}
\left\{
\begin{aligned}
\tilde{e}_{\res}(\tauuntilde) &\simeq \frac{\tilde{\tau}_e}{\tauuntilde}
\quad \mbox{for} \quad \tauuntilde \lesssim \tilde{\tau}^{\crit}_N \\
\tilde{\tau}^{\crit}_N &= \frac{N-1}{2}
\end{aligned}
\right. \;,
\end{equation}
where $\tilde{\tau}_e$ is a dimensionless constant of order one, and the
critical time $\tilde{\tau}^{\crit}_N$ grows linearly with the system
size~\cite{Mbeng_PhDThesis2019New,Arezzo_inpreparation}. Since, by
Eq.~\eqref{eq:eres_equal}, the physical and reference problems have the
same residual energy, and, by Eq.~\eqref{eq:bare_physical_time_relation},
their durations differ by the factor $\hbar/(2J^*)$,
Eq.~\eqref{eq:bare_residual_scaling} becomes, in terms of the physical
unitary-stroke duration,
\begin{equation} \label{eq:physical_residual_scaling}
\eres(\tauun) \simeq \frac{\tau_e}{\tauun}
\quad \mbox{for} \quad \tauun \lesssim \tau^{\crit}_N \;,
\end{equation}
with
\begin{equation} \label{eq:tau_e_crit_rescaling}
\left\{
\begin{aligned}
\tau_e &= \frac{\hbar}{2J^*}\,\tilde{\tau}_e
        = \frac{\hbar(g_c+J_h)}{2g_cJ_h}\,\tilde{\tau}_e \\
\tau^{\crit}_N &= \frac{\hbar}{2J^*}\,\tilde{\tau}^{\crit}_N
        = \frac{\hbar(g_c+J_h)}{4g_cJ_h}\,(N-1)
\end{aligned}
\right. \;,
\end{equation}
which are Eqs.~\eqref{eqn:eres_opt} and~\eqref{eq:tau_crit}. For our
endpoints, $\hbar/(2J^*)\approx3.03$ in units of $\hbar/(k_BT_h)$, and the
residual energy of the optimized protocol in
Fig.~\ref{fig:cycle_work_comparison}(b) gives $\tilde{\tau}_e\approx1$.

The same rescaling applies to the threshold time of
Eq.~\eqref{eq:tau_star_fit}, $\tau^{\thrs}_N=\tau_1N+\tau_0$, which
corresponds to $\tilde{\tau}^{\thrs}_N=\tilde{\tau}_1N+\tilde{\tau}_0$ with
$\tilde{\tau}_{0,1}=(2J^*/\hbar)\,\tau_{0,1}$, i.e.,
$\tilde{\tau}_1\approx0.54$ and $\tilde{\tau}_0\approx1.7$. The slope
$\tilde{\tau}_1$ is close to the slope $1/2$ of $\tilde{\tau}^{\crit}_N$,
consistently with the fast crossover between the two times. 
In the sharp-crossover approximation of
Eq.~\eqref{eqn:epsilon_res_tau}, the residual energy is set to zero for
$\tauun\ge\tau^{\thrs}_N$.
 
\section{Kibble--Zurek scaling of the linear schedule}
\label{app:kz}

For the linear schedule $s(t)=t/\tauun$, the forward Hamiltonian in each
positive-momentum sector reads, see App.~\ref{app:thermoising},
\begin{equation}
H_k(t) = 2\Big[\big(g_c(1-s) - J_h s\cos k\big)\tau_k^z
+ J_h s\sin k\,\tau_k^x\Big] \;.
\end{equation}
Writing $H_k=\mathbf{h}_k(t)\cdot\boldsymbol{\tau}_k$, with $\boldsymbol{\tau}_k=(\tau^x_k,\tau^z_k)$, the effective field
$\mathbf{h}_k$ moves along a straight line in the $(x,z)$ plane with
constant speed $|\dot{\mathbf{h}}_k| = 2\bar\epsilon_k/\tauun$, where $\bar{\epsilon}_k=\sqrt{g_c^2+J_h^2+2g_cJ_h\cos k}$.
The distance of this line from the origin is
$d_k=2g_cJ_h\sin k/\bar{\epsilon}_k$.
The closest point is reached at
\begin{equation}
    s_{k,{\rm \min}}= \frac{g_c(g_c+J_h\cos k)}{
g_c^2+J_h^2+2g_cJ_h\cos k} \;,
\end{equation}
lying within the physical interval $0<s<1$ when
$\cos k>-\min(g_c/J_h,J_h/g_c)$; in that case, $d_k$ is half the minimum gap of the sector.
For the small-$k$ modes relevant to Kibble--Zurek scaling,
$s_{k,{\rm \min}}\to s^*$ as $k\to0$, where $s^*$ locates the critical point.
For these small-$k$ modes and sufficiently slow ramps, the endpoints lie far from the avoided crossing on the Landau--Zener scale. The finite-time passage can then be approximated by an asymptotic Landau--Zener sweep, giving
\begin{equation} \label{eq:lz_pk}
p_k \simeq \exp\!\left(-\frac{\pi d_k^2}{\hbar|\dot{\mathbf{h}}_k|}\right)
= \exp\!\left(-\frac{2\pi\tauun}{\hbar}\,
\frac{g_c^2J_h^2\sin^2 k}{\bar\epsilon_k^{3}}\right) \;,
\end{equation}
which is the generalization to our path of the mode-resolved
Landau--Zener analysis of the transverse-field Ising chain of
Refs.~\cite{Dziarmaga_PRL2005,mbeng2024quantum}.
 
The leading contribution for slow ramps comes from small momenta. Using Eq.~\eqref{eq:forward_residual_fidelities} and defining $p_k = 1-F_k^f$, we approximate the small-momentum excitation probabilities by Eq.~\eqref{eq:lz_pk} and
replace the sum over the $N/2$ positive momenta by an integral, obtaining
\begin{equation}
\eres^{\lin} = \frac{4}{N}\sum_{k>0}p_k
\approx \frac{2}{\pi}\int_0^{\pi/2} dk\; p_k \;.
\label{eq:elinres}
\end{equation}
Here, $\pi/2$ is a convenient cutoff for the small-momentum contribution, rather than the boundary for the existence of an interior minimum gap. Its precise value does not affect the leading scaling when $k_{\mathrm{KZ}}\ll 1$.
In this regime,
$\sin k\simeq k$ and $\bar\epsilon_k\simeq g_c+J_h$. Equation~\eqref{eq:lz_pk}
then becomes a Gaussian, $p_k\simeq e^{-k^2/k_{\rm KZ}^2}$, with
\begin{equation}
k_{\rm KZ}^{-2} = \frac{2\pi J^{*2}}{\hbar\,(g_c+J_h)}\,\tauun \;,
\qquad J^* = \frac{g_cJ_h}{g_c+J_h} \;,
\end{equation}
and the integral gives
$\eres^{\lin}\simeq k_{\rm KZ}/\sqrt{\pi}$, i.e.,
Eq.~\eqref{eq:kz_linear}. This approximation requires $k_{\rm KZ}\ll1$
and, for a finite chain, $k_{\rm KZ}\gg\pi/N$, i.e.,
\begin{equation}
\label{eqn:adiabaticthreshold}
\frac{\hbar\,(g_c+J_h)}{2\pi J^{*2}} \ll \tauun \ll
\frac{\hbar\,(g_c+J_h)}{2\pi^3 J^{*2}}\,N^2 \;.
\end{equation}
For our parameters this window is $6\ll\tauun\ll 5\times10^{2}$, in units of $\hbar/(k_BT_h)$ at
$N=30$. The integral in Eq.~\eqref{eq:elinres} differs from
Eq.~\eqref{eq:kz_linear} by less than $4\%$ already at $\tauun=12\,\hbar/(k_BT_h)$.
When $k_{\rm KZ}<\pi/N$, even the mode with the smallest gap, $k_1=\pi/N$, is crossed adiabatically
and the residual energy drops, which gives the adiabatic threshold set by the upper bound of Eq.~\eqref{eqn:adiabaticthreshold}.
 
\bibliography{biblio}
\end{document}